\documentclass{aa}  

\usepackage{graphicx}
\usepackage{txfonts}
\begin{document} 

   \title{Central engine of GRB170817A: Neutron star versus Kerr black hole based on multimessenger calorimetry and event timing}
   
   \titlerunning{Central engine of GRB170817A}

   \subtitle{} 

   \author{Maurice H.P.M. van Putten
          \inst{1}
          \and
          Massimo Della Valle\inst{2} 
          }
   \institute{Department of Physics and Astronomy, 
              Sejong University, 209 Neungdong-ro, 
              Gwangin-gu, Seoul 05006, Republic of Korea\\
              \email{mvp@sejong.ac.kr}
         \and
             INAF- Osservatorio Astronomico di Capodimonte, Salita Moiariello 16, 80131 Napoli, Italy and Department of Physics, Ariel University, Ariel, Israel\\
             }
   \date{}

 
  \abstract
  {   
  LIGO-Virgo-KAGRA observations may identify the remnant of compact binary coalescence and core-collapse supernovae associated with gamma-ray bursts. The multimessenger event GW170817-GRB170817A appears ripe for this purpose thanks to its fortuitous close proximity at 40\,Mpc. Its post-merger emission, ${\cal E}_{GW}$, in a descending chirp can potentially break the degeneracy in spin-down of a neutron star or black hole remnant by the relatively large energy reservoir in the angular momentum, $E_J$, of the latter according to the Kerr metric.}
  {The complex merger sequence of GW170817 is probed for the central engine of GRB170817A by multimessenger calorimetry and event timing.}
  {We used {model-agnostic} spectrograms with equal sensitivity to ascending and descending chirps {generated by time-symmetric butterfly matched filtering.} 
  The sensitivity was {calibrated} by response curves generated by software injection experiments, covering a broad range in energies and timescales. The statistical significance for candidate emission from the central engine of GRB170817A is expressed by probabilities of false alarm (PFA; type I errors) derived from an event-timing analysis.
  {Probability density functions (PDF) were derived for start-time $t_s$}, identified via high-resolution image analyses of the available spectrograms.
  For merged (H1,L1)-spectrograms of the LIGO detectors, a PFA $p_1$ derives from causality 
  {in $t_s$ given GW170817-GRB17081A} (contextual). 
  A statistically independent confirmation is presented in individual H1 and L1 analyses, quantified by a second PFA $p_2$ of consistency {in their respective observations of $t_s$} (acontextual). 
  A combined PFA derives from their product since the mean and (respectively) the difference in timing are statistically independent.}
  {Applied to GW170817-GRB170817A, {PFAs of event timing in $t_s$} produce $p_1=8.3\times 10^{-4}$ and $p_2=4.9\times 10^{-5}$ {of a post-merger output} ${\cal E}_{GW}\simeq 3.5\%M_\odot c^2$ ($p_1p_2=4.1\times 10^{-8}$, equivalent $Z$-score 5.48). ${\cal E}_{GW}$ exceeds $E_J$ of the hyper-massive neutron star in the immediate aftermath of GW170817, yet it is consistent with $E_J$ rejuvenated in gravitational collapse to a Kerr black hole. Similar emission may be expected from energetic core-collapse supernovae producing black holes of interest to upcoming observational runs by LIGO-Virgo-KAGRA.
  }
  {}
  
   \maketitle

\section{Introduction}

Gamma-ray bursts represent the most relativistic transient events in the sky. Discovered serendipitously \citep{kle73}, 
they are now known to have an astronomical origin in the end point of massive stars, broadly classified as short duration
events (SGRBs), some with extended emission (SGRBEEs), and long-duration events \citep[LGRBs, reviewed in e.g.,][]{van14b}.
The end of life of relatively massive stars is believed to be the origin of neutron stars and stellar mass black holes by gravitational collapse when their nuclear burning phase ceases. Neutron stars can be found in supernova remnants harboring pulsars, as single objects such as Vela and the Crab \citep{hew70}, but also in binaries, notably the Hulse-Taylor system PSR1913+16 \citep{hul75} 
and the short-period binary PSR J0737-3039 \citep{bur03}.

While pulsar emission gives unambiguous evidence of spinning neutron stars, identifying stellar mass black holes (at similar levels of rigor) appears more challenging, despite their common astronomical origin in core-collapse of massive stars. Following the seminal discovery of cosmological redshift of gamma-ray bursts in optical follow-up observations \citep[e.g.,][]{blo01} to an X-ray afterglow to GRB970228 by {\em BeppoSAX} \citep{cos97}, LGRBs have been identified thanks to core-collapse supernovae  -- energetic events of type Ib/c from relatively more massive stars \citep{gal98,hjo03,sta03,mat03,mod06,gue07,kel08}. On the other hand, GRBs originating in compact binary coalescence \citep{pac86} increasingly appear to be of the short-duration variety, as notably demonstrated by GW170817-GRB170817A \citep{abb17c}; however, this may be extended to include long-duration events, depending on black hole spin \citep{van03} or progenitor binary evolution \citep[e.g.,][]{rue21}.

The astronomical origin of GRBs identified with the end-point of stellar evolution leaves the central engine producing the ultra-relativistic outflows responsible for their non-thermal gamma-ray emission to be either a strongly magnetized neutron star (magnetar) or black hole \citep[e.g.][]{pir04,pir19a,pir19b,nak20}. In contrast to neutron stars, electromagnetic observations alone appear to be insufficient to conclusively distinguish between the two, despite a half-century of multi-wavelength observations covering GRBs by their prompt gamma-ray emission and afterglows in
X-rays and optical-radio.

The recent advance of the {LIGO-Virgo-KAGRA (LVK}) detectors offers unprecedented and independent opportunity to advance this conundrum. Quite generally, an astrophysical black hole central engine powering an extreme transient event will be exposed to surrounding high-density matter. Gravitational radiation thereby produced is expected to be dominant over MeV-neutrinos and electromagnetic radiation when powered by it ample angular momentum energy reservoir in its angular momentum \citep{van03} which, according to the Kerr metric, readily exceeds the same of a neutron star by an order of magnitude. True calorimetry hereby promises the break the degeneracy of neutron stars and black holes as the central engine, upon including total energy output in gravitational radiation (GW-calorimetry). 

Observational opportunities may be found in compact binary mergers involving a neutron star \citep{abb17c} and core-collapse supernovae, especially those associated with LGRBs \citep{lsc18,van19c}. At a distance $D\simeq 40$Mpc \citep{cou17,can18}, GW170817 presents a fortuitously nearby event offering a unique prospect to witness the formation of stellar mass black holes in gravitational collapse, directly or delayed, of an intermediate hypermassive neutron star formed in the immediate aftermath of this double neutron star merger \citep[e.g.,][]{mur20} -- in addition to the new opportunities for probing physical properties of the neutron star progenitors \citep{abb18a,dra18,dep19,bau19}. 

In association with GRB170817A, the precursor emission GW170817 makes it feasible to achieve a key objective of multimessenger observations \citep[][]{ace07}. An earlier demonstration of significance in timing across different observational channels is found in SN1987A, with essentially coincident arrival times of an MeV-neutrino burst detected independently by Kamiokande II and IMB, and  alongside an optical light curve of SN1987A in the satellite galaxy LMC. For GW170817, independent pickup of the merger signal GW170817 by H1 and L1 is likewise coincident (within the light travel time of 10\,ms between H1 and L1), followed by GRB170817A in NGC4993. At the distance of 40\,Mpc, however, any further MeV-neutrino emission is a priori undetectable. Quite generally, timing coincidences carry observational significance based on a uniform prior on event timing on scales less than a Hubble time, subject only to instrumental constraints.

\begin{figure}
\centerline{\includegraphics[scale=0.14]{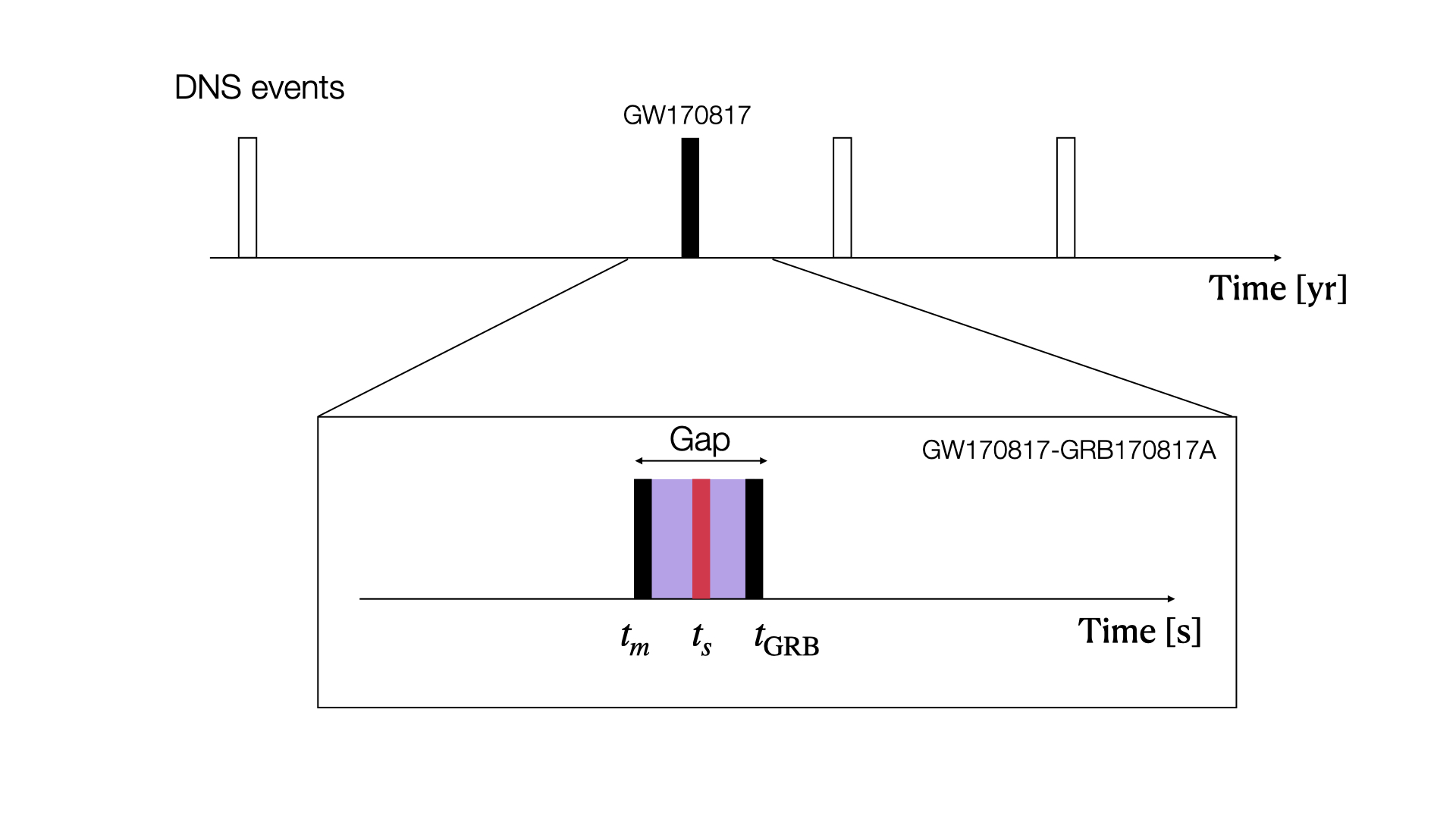}}
\caption{
Schematic of event timing in the first double neutron star merger detected (top): the multimessenger event GW170817 at merger time $t_m$ with accompanying GRB170817A at $t_{\rm GRB}=t_m+t_g$ following a gap $t_g\simeq 1.7\,s$ (bottom). The central engine of GRB170817A formed in this gap by causality, whereby $t_m<t_s<t_{GRB}$ for the start-time of accompanying gravitational radiation. This gap condition poses a {\em discrete event time} problem for the probability $p$ of false alarm (PFA) for a candidate detection to be in the 1.7\,s gap between GW1709817-GRB170817A: $p=t_g/T$ marked by the (unique) global maximum of an indicator function over an observational time $T$ given the uniform prior on astrophysical event times (Appendix A). 
When sufficiently large, the energy output ${\cal E}_{GW}$ marked by $t_s$ can break the degeneracy between neutron stars and black holes.
}
\label{fig_C1}
\label{figC1}
\end{figure}

Gravitational wave observations of compact binary mergers is realized at observational sensitivity close to the limit defined by strain-noise amplitude of the detectors \citep[detector limit;][]{abb21}. This limit is defined by the theory of ideal matched filtering in the face of finite detector noise exploiting phase-coherence in correlation with model gravitational-wave templates over a large number of wave periods \citep[e.g.,][]{abb20}. Ideally, un-modeled searches are pursued at similar sensitivities, even when phase-coherence is limited to intermediate time-scales (between the period of the wave and the total duration of the event). For transient emission featuring ascending or descending frequencies in gravitational-wave emission, broadband spectrograms provide a suitable starting point to the search for transient emission by image analysis. High sensitivity spectrograms can be realized by butterfly filtering over a dense bank of time-symmetric templates of intermediate duration $\tau$ with sensitivity exceeding that of Fourier-based spectrograms by over an order of magnitude \citep{van14a}. 
It and high-resolution mage analysis can be accelerated on modern {graphic processor units} (GPUs) 
with {high bandwidth memory} (Appendix D). 
This opens a window to probing the central engine of GRB170817A by GW-calorimetry at sensitivities on par with the pre-cursor GW170817.

The statistical significance of a candidate gravitational-wave event derives from the probability of false alarm (PFA), that is, Type I errors assuming the null-hypothesis of no gravitational waves passing by the LIGO detectors. Multimessenger observations involving the H1 and L1 detectors suggest exploiting both energy and event times $t_k=t_k(t)$: a Boolean stochastic variable over $\{0,1\}$ as a function of real time $t$ indicating an event $k$ at some specific instant $t_s$ \citep[e.g.,][]{bro04,don16} (Fig. \ref{figC1}). For a candidate emission feature, PFAs may derive from consistency conditions applied to a probability density (PDF) of $t_s$. Consistency in timing can be contextual or acontextual. Quite generally, this can be approached with and without priors from other observational channels, shown here in merged and individual H1 and L1 analyses, respectively.

Event timing of a candidate post-merger emission feature to GW170817 associated with the central engine of GRB170817A is subject to the gap condition (Fig. \ref{figC1}):
\begin{eqnarray}
C_1: t_s\,\epsilon \,G
\label{EQN_C1}
,\end{eqnarray}
given the interval $G=\left[t_m,t_m+t_g\right]$ as an observational prior on its start-time, $t_s$, by causality; $C_1$ hereby introduces a Boolean-valued statistic, $t_k(t)=1$ $\left(t\,\epsilon\,G\right),$ and 0 otherwise. 
Given the uniform prior on astrophysical event times, such a discrete event time carries a PFA equal to $t_g/T$ over an observation of duration $T$ as a conventional probability. {As elucidated in Appendix A, this is analogous to the odds of winning a bet in roulette provided that the measurement of $t_s$ is  sufficiently precise.}

Combining PFAs in one-class classification schemes \citep[e.g.][]{sim17} is common in image analysis, here spectrograms in \S3-4. Quite generally, we differentiate between continuous or discrete distributions \citep[e.g.][]{blo06}. The product of two $p$-values of continuous distributions, for instance, signal amplitude, is not a $p$-value, but can be merged by the celebrated Fisher's combined probability test \citep{fis32,fis48,sim17,lyo18}. For a pair of statistically independent $p$-values $p_i$ $(i=1,2)$, Fisher's method defines an equivalent $p$-value by equivalent information content $I_i=-\log p_i$ according to $\chi_F^2=-2\log P$, where $P=p_1\times p_2$ and $\chi^2_F$ denotes the $\chi^2$ distribution with $F=4$ degrees of freedom. The merged $p$-value derives from the cumulative density function of $\chi^2_F$.  Fisher's method has been widely used and indeed led to a number of potential improvements \citep{whi05,hea17}. When $p$-values $p_1$ and $p_2$ are small, $p=P\left(1-\log P\right)$ gives a merged $p$-value that is effectively equivalent to Fisher's \citep{the04,lyo18}. This is not the case in the present discrete event timing analysis of $C_1$, where the PFA derives as a conventional probability for (\ref{EQN_C1}) that can be merged as such with other PFAs \citep[e.g.,][]{the04}.

In the next section, we zoom in on the complex merger sequence of the double neutron star merger GW170817 with the aim of identifying the central engine -- a hyper-massive neutron star or rotating black hole -- of the associated GRB170817A by GW-calorimetry and event timing, previously introduced in \cite{van19a,van19b}. These works are hereafter referred to Paper I and Paper II, respectively.

To this end, we present two statistically independent analysis of H1 and L1 data by model-agnostic spectrograms.
Observational results are compared and combined based on high-resolution PDF($t_s$)s realized by modern exascale heterogeneous computing (Appendix D). In particular, a joint PFA from event timing, factored over PFAs derived from PDF($t_s)$s of each, provides a significantly enhanced level of confidence in the observations.

To start (\S3), we revisit and extend our analysis of merged (H1,L1)-spectrograms in Paper I and II with (a) a one-parameter family of response curves calibrated by signal injection experiments and (b) PDF$(t_s)$ generated over an extended foreground derived from a large number of small time-slides between H1 and L1, satisfying (c) a condition of
extremal clustering robust against false positives from data anomalies. Statistical significance is expressed by PFA$_1$ from event timing by causality in the context of GW170817-GRB170817A ($C_1$) (Appendix A).

Next (\S4), we present a statistically independent analysis of H1 and L1 individually with accompanying PDF$(t_s)$. Statistical significance is expressed by PFA$_2$ of consistency in timing, $t_s$, by cross-correlating their respective PDF$(t_s)$s ($C_2$). 
In \S5, we {discuss} our results and the observational consequences for the central engine of GRB170817A with joint PFA factored over PFA$_1$ and PFA$_2$ based on statistical independence of mean and difference in event time observations, providing a key advance over previous analysis.
We summarize our conclusions (\S6) with an outlook (\S7) on upcoming opportunities to probe merger sequences involving neutron stars and core-collapse supernovae in the Local Universe (\S6).

\section{Rejuvenation in gravitational collapse}

The double neutron star (DNS) merger sequence of GW170817 is complex, involving a hyper-massive neutron star (HNS) formed in the immediate aftermath \citep{abb17c,dai19,dep19b,mur20} with the possibility of (delayed) gravitational collapse to a rotating black hole \citep[BH;][]{aku20}:
\begin{eqnarray}
\begin{array}{lll}
 {\rm 
 \mbox{DNS} \rightarrow \mbox{HNS} \rightarrow \mbox{BH}+\mbox{GRB170817A} + \mbox{AT2017gfo}+\cdots
 , }
\end{array}
\label{EQN_MS}
\end{eqnarray}
with the ellipses referring to additional post-merger emission in MeV-neutrinos and gravitational radiation.
In a continuing gravitational collapse -- direct or delayed -- a black hole will form surrounded by a high-density disk or torus from debris and dynamical mass-ejecta from the merger prior \citep[e.g.,][]{ros99,bai17,cio20}. In Eq. \ref{EQN_MS} the GW170817-GRB170817A association appears quite secure with independent $p$-values from continuous random variables: temporal ($P_{\mbox{\tiny temp}}=5\times 10^{-6}$) and spatial ($P_{\mbox{\tiny spat}}=1\times 10^{-2}$) consistency in the LIGO H1 and L1-detectors, on the one hand, and GRB170817A detected by {\em Fermi} Gamma-ray Burst Monitor (GBM) \citep{abb17c}. 

In Eq. \ref{EQN_MS}, the potential for breaking aforementioned degeneracy between a neutron star or black hole remnant derives, in part, from  a rejuvenation of the energy reservoir in angular momentum, $E_J$, in the process of gravitational collapse. By conservation of specific angular momentum $a/M=\sin\lambda$ in prompt gravitational collapse of the HNS with spin-energy $E_J^-\simeq  (1/2)J\Omega$, 
$E_J^+= 2Mc^2\sin^2(\lambda/4)$ of the black hole \citep{ker63} produces:
\begin{eqnarray}
\frac{E_J^+}{E_J^-}  = \frac{(R/R_g)^2}{10\cos^2(\lambda/4)\cos^2(\lambda /2)} \simeq 2.5 \left(\frac{R}{18\,{\rm km}}\right)^2.
\label{EQN_ratio}
\end{eqnarray}
Here, the right hand size refers to GW170817, using $J=I\Omega$ for the moment of inertia $I\simeq (2/5)MR^2$ of the HNS with gravitational radius $R_g=GM/c^2$, radius $R$ and angular velocity $\Omega$ with Newton's constant $G$ and the velocity of light $c$. Further enhancement in $a/M$ might derive from hyper-accretion \citep[e.g.,][]{bar70,lev13}, although this does not appear to be called for in GW170817. 
Therefore, in the case of the gravitational collapse of a $\sim 3M_\odot$ hyper-massive neutron star,  $E_J$ increases by Eq. \ref{EQN_ratio}, setting the initial condition for potentially significant post-merger emission from a black hole-torus system.
 
\subsection{Calorimetry in gravitational radiation}

Indeed, rejuvenation Eq. \ref{EQN_ratio} can safely account for Extended Emission feature in (H1,L1)-spectrogram (Fig. \ref{fig_tws}) with (Paper II): 
\begin{eqnarray}
{\cal E}_{gw}=(3.5\pm1)\%M_\odot c^2.
\label{EQN_2}
\end{eqnarray}
Notably, (\ref{EQN_2}) breaks the degeneracy between neutron stars and black holes, since ${\cal E}_{gw}\simeq 6.3\times 10^{52}$erg exceeds the limit of rotational energy of the hypermassive neutron star progenitor, even more so after the initial one 
second of neutrino cooling and internal dissipation leading to a uniform rotator - a supermassive neutron star \citep{ben21}. 
Further by the observed gravitational-wave frequencies $f_{gw}\lesssim 700\,$Hz (Fig. \ref{figEE}) would imply a spin frequency well-below the Keplerian frequency limits \citep{hae11} for quadrupole emission at twice its spin-frequency, limiting its rotational energy to: 
\begin{eqnarray} 
E_J^- = \frac{\pi^2}{5}  f_{gw}^2 M R^2\lesssim 1.6\times 10^{52} \left(\frac{M}{2.5M_\odot}\right) \left(\frac{R}{18\,\mbox{km}}\right)^2\,\mbox{erg}.
\label{EQN_2b}
\end{eqnarray}
Thus, Eq. \ref{EQN_2} exceeds Eq. \ref{EQN_2b} by a factor greater than 4, defined by maximal values of mass $M$ and radius $R$. Yet, Eq. \ref{EQN_2} is in quantitative agreement with the output of black hole spin-down against a non-axisymmetric high-density disk or torus (Paper II). 

Accordingly, GW-calorimetry hints at continuing gravitational collapse to a black hole in 
Eq. \ref{EQN_MS}. 
In contrast, calorimetry in the (post-merger) electromagnetic observation of GRB170817A and kilonova AT 2017gfo \citep{con17,sav17,moo18a,moo18b,asc20} shows a combined energy output limited to
\begin{eqnarray}
{\cal E}_{EM}\simeq 0.5\%M_\odot c^2.
\label{EQN_EM1}
\end{eqnarray}
This output falls woefully short to break the degeneracy between a neutron star of black hole remnant in the chronicle Eq. \ref{EQN_MS}, prompting the need for GW-calorimetry outlined above.

Even so, electromagnetic observations do provide us with crucial timing information. Post-merger rejuvenation in Eq. \ref{EQN_MS} appears delayed by constraints on the lifetime of the hypermassive neutron star \citep{poo18,gil19,rad18,luc19}, inferred from a ``blue" component  of the associated kilonova AT 2017gfo \citep{sma17,pia17} (but see \cite{ren19,lu19,pir19a} for a possibly long-lived supermassive neutron star remnant). Reviewed in \cite{mur20}, Fig. \ref{fig_tws} points to a lifetime of $\lesssim\,1$\,s based on various independent studies 
\citep{gra17,got18,nak18,met18,xie18,gil19,laz20,ham20}. 

\subsection{Sensitivity requirements}

Performing GW calorimetry Eq. \ref{EQN_2} to post-merger emission requires sensitivity on par with the CBC search for GW170817. Gravitational radiation from non-axisymmetric disks or tori swept up by a rotating black hole lack the phase-coherence over long duration, normally exploited in model-dependent searches applicable to compact binary coalescence (CBC). 

Existing un-modeled searches by power-excess methods \citep{abb17c,abb19a,abb19b} fall short by a threshold for detection of ${\cal E}_{th,gw}\gtrsim 6.5M_\odot c^2$ \citep{sun19}. This represents a sensitivity to post-merger signals of 0.3\% compared to ${\cal E}_{gw}\simeq2.5\%M_\odot c^2$ in GW170817 itself \citep{abb17c}, {a priori in an excluded zone of the parameter space by the total progenitor mass-energy
$E_{DNS}\lesssim 3M_\odot c^2$ of GW170817. }

Here, we apply butterfly matched filtering which aims to capture signals with frequencies slowly wandering in time. Using a dense filter bank of time-symmetric templates, it realizes equal sensitivity to ascending and descending branches relevant to the present study of merger and post-merger emission to Eq. \ref{EQN_MS}, squarely in the allowed zone
${\cal E}_{GW}<E_{DNS}$ of any post-merger radiation.

As a pass filter for frequencies with finite time rate-of-change in frequency, constant frequency signals are relatively suppressed \citep{van16}, while stochastic signals can be detected at sensitivity far beyond the sensitivity of conventional Fourier analysis \citep{van14a}. A precise characterization for the purpose of the present study follows and is based on detailed signal injection experiments.

\section{Analysis of merged (H1,L1)-spectrograms}

In this section, we consider (H1,L1)-spectrograms merged by frequency coincidences, generated by butterfly matched filtering (Paper I and II). 
Merged (H1,L1)-spectrograms were considered for the following  steps (Appendix C): 1)  the sensitivity is the same for ascending and descending chirps by time-symmetry of matched filtering templates (this step is un-modeled). 2) parameter estimation is performed by an image scan, quantifying features by a goodness of fit over a family of descending chirps using $\chi$-image analysis (detailed in Papers I and II).

Extending previous work, we analyze merged spectrograms by application of surrogate time-slides of the following type: 
{\bf ($S_0$)} {foreground} in conventional analysis of merged (H1,L1)-spectrograms with zero time-slide, $\Delta t =0$;
{\bf ($S_1$)} {extended foreground} over small time-slides $\Delta t$ less than the intermediate duration $\tau$ of our templates;
{\bf ($S_2$)} {background} from time-slides $\Delta t$ greater than $\tau$. We note that $\tau$ is considerably greater than the light travel time between H1 and L1.

We extract PDF($t_s)$ from extended foreground ($S_1$). This is a principle extension to previous work, restricted to {foreground} $(S_0)$, that is, zero time-slide. 
A high-resolution PDF($t_s)$s produced by heterogeneous computing (Appendix D) permit accurate parameter estimation, rigorous application of event timing (\S3.2, Appendix A), robustness against false positives by extremal clustering (\S3.3), and enable a direct comparison and combination of results between statistically independent analyses (\S4).

\begin{figure*}
\centerline{
\includegraphics[scale=0.19]{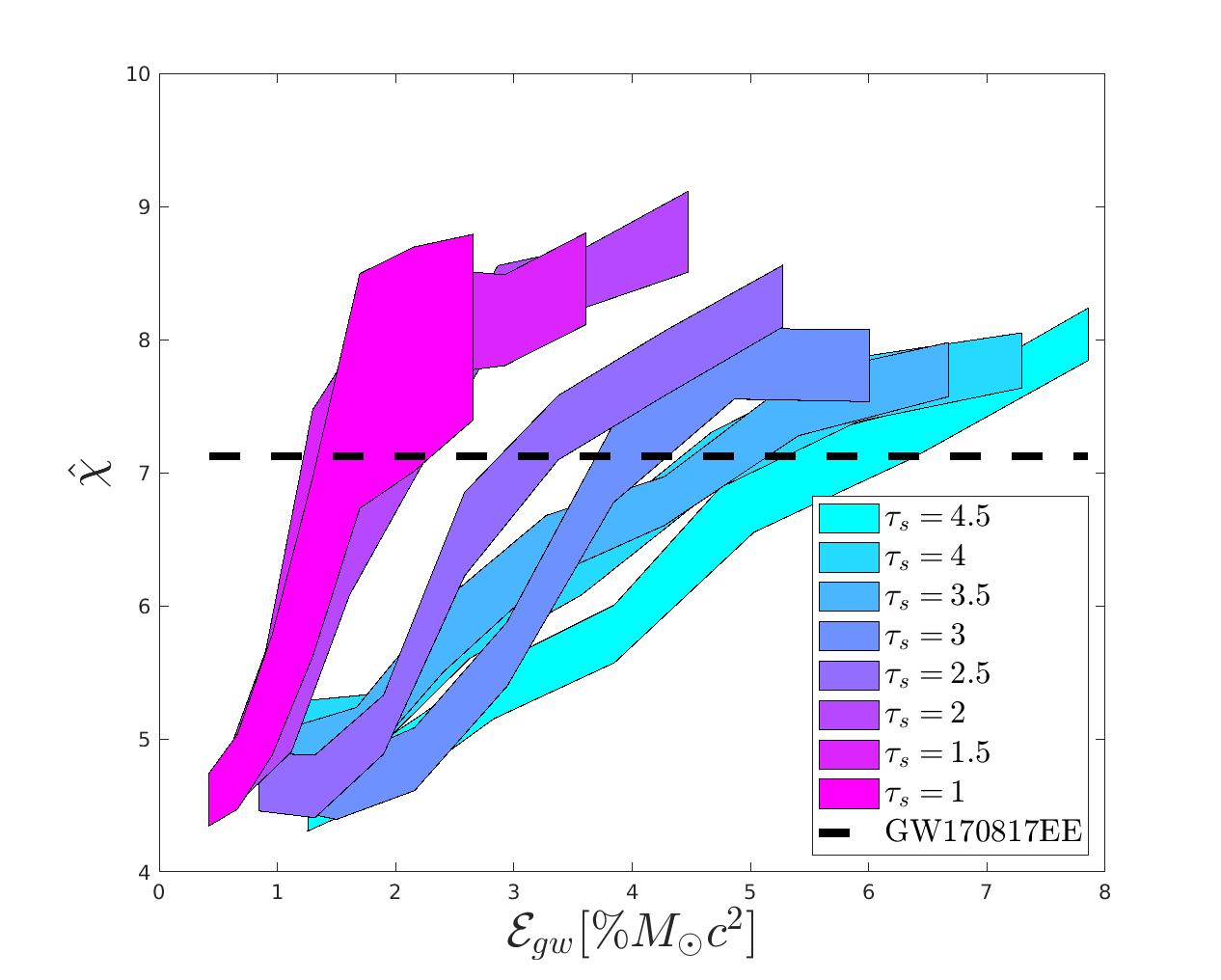}
\includegraphics[scale=0.19]{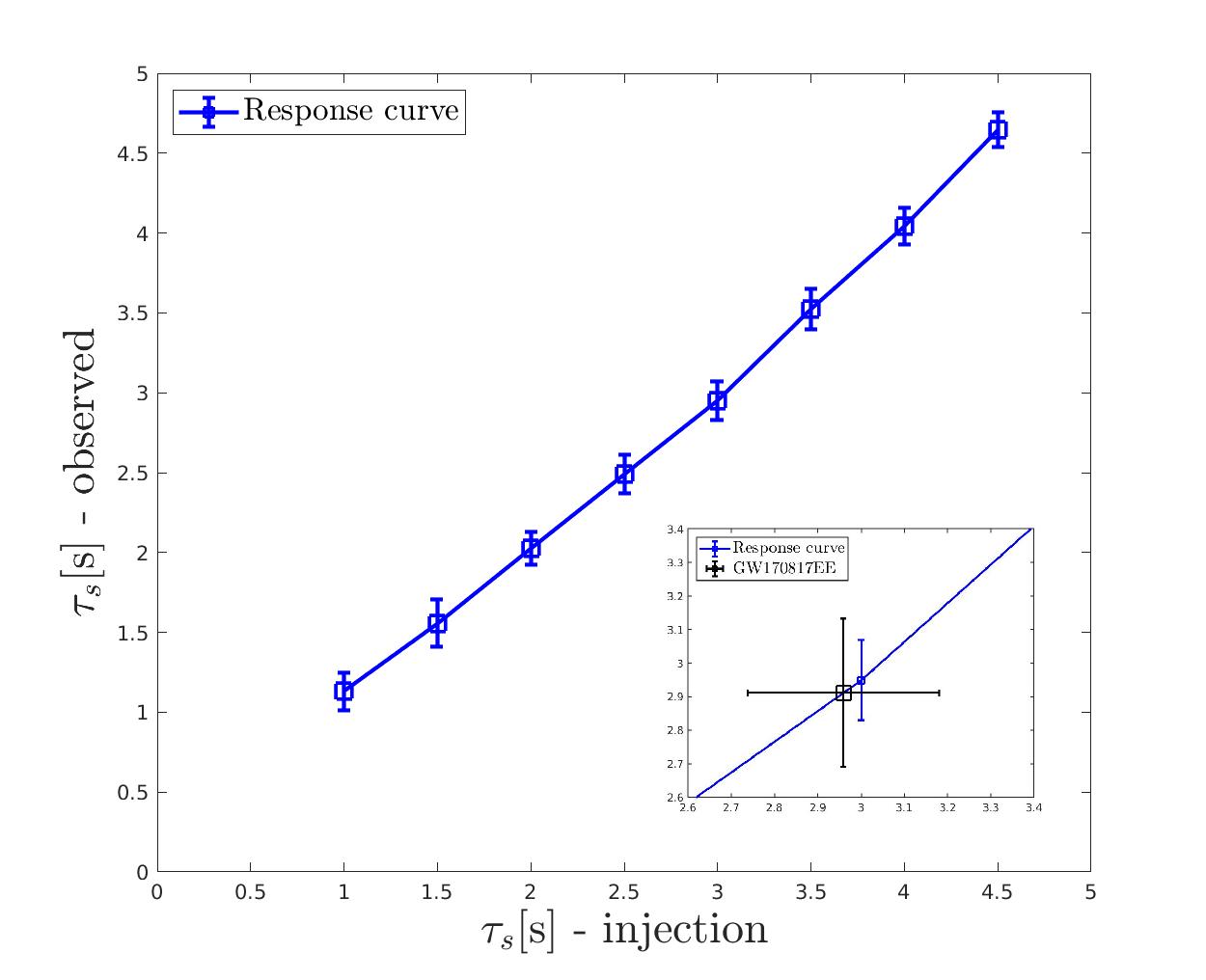}
}
\centerline{\mbox{~}\includegraphics[scale=0.203]{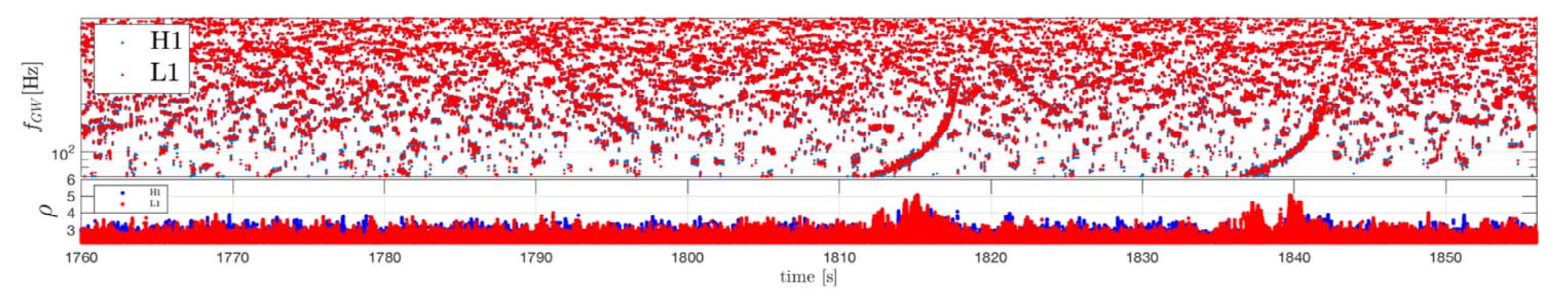}}
\caption{
Eight response curves of the two-level pipeline by $\chi$-image analysis of merged (H1,L1)-spectrograms produced by butterfly matched filtering using a total of 56 signal injection experiments on LIGO O2 data covering GW170817 (top-left panel). These model injections are defined by Eq. \ref{EQN_c}. 
The dashed line is the mean of $\hat{\chi}$-peaks averaged over (tiny) time-sides $\Delta t$ within the light-travel time of 10\,ms between H1 and L1 identified with extended emission to GW170817EE in a post-merger descending branch (lower panels). Precision realized is shown by parameter recovery (top-right panel). The parameter ``field-of-view" covered for descending chirps covers energies $0<{\cal E}_{gw}<8\%M_\odot c^2$ (7 steps) and timescales of descent $\tau_s$ (eight steps, in units of seconds) with start frequency $f_s=680$\,Hz for a model source distance $D=40\,$Mpc given by GW170817. These injection signals are in the shot-noise dominated regime of the LIGO detectors, where noise increases with frequency and hence sensitivity decreases with $\tau_s$. 
Inserted is a plot of observed $\tau_s$ of post-merger emission to GW170817 (black square), overlaid to the calibration curve (blue curve) to infer uncertainty (along abscissa) from observed scatter (along the ordinate).
Sample of a single injection of a combined merger and post-merger signal (1810-1825\,s) in the snippet of H1L1-data containing GW170817 ($t_m=1842.43\,$s) (reprinted from Paper II), shown in the lower panel.
The present 56 injections Eq. \ref{EQN_c} vary over energy ${\cal E}_{GW}$ and time-scale of descent $\tau_s$, keeping merger signal injection fixed at the parameters of GW170817.} 
\label{figRes}
\end{figure*}

\subsection{Response curves to descending branches} 
Figure \ref{figRes} shows calibrated response curves 
computed by model signal injections. In what follows, all injections are a model merger chirp followed by a post-merger descending chirp, where the first is fixed according to the parameters of GW170817 (Paper II, Fig. \ref{figRes}). 

Our descending chirp follows a fit to the post-merger feature attributed to radiating $E_J$ of the compact remnant, distinct from an ascending chirp in binary coalescence, given by (Paper I and II):
\begin{eqnarray}
f(t)=(f_s-f_0)e^{-(t-t_s)/\tau_s}+f_0.
\label{EQN_c}
\end{eqnarray}
These curves introduce four parameters: a start frequency, $f_s$, late-time frequency, $f_0$, in addition to aforementioned $(t_s,\tau_s)$ used in the $\chi$-image analysis of spectrograms. Stable estimates derive in goodness of fit over data-segments of $\Delta t = 7\,$s, moving over small time-steps of 33\,ms (supplementary data, Paper I). 

Our response curves are generated by injection experiments cover the two-parameter space of energies and $\tau_s$,
\begin{eqnarray}
{\cal E}_{GW}=\left( 0.5-8\right)\%M_\odot c^2, ~~\tau_s=\left(1-4.5\right)\,\mbox{s},
\label{EQN_FOV}
\end{eqnarray}
using a source distance of $D=40$\,Mpc of GW170817  \citep{can18}, $f_s=680$\,Hz, and $f_0=98$\,Hz.

We identify Eq. \ref{EQN_c} in merged (H1,L1)-spectrograms by $\chi$-image analysis parameterized by $(t_s,\tau_s,f_s,f_0)$ 
at a combined resolution of 0.5\,MHz ($2^{14}$ steps in ($\tau_s,f_s,f_0$) at 30\,Hz in $t_s$) in a high-frequency scan over $10^9$ parameter values over the 2048\,s frame of H1L1-data sampled at 4096Hz. Such is applied following a time-slides applied to H1 and L1 (prior to generating spectrograms).

Figure \ref{figRes} shows results produced by 56 injections over ${\cal E}_{gw}$ in seven steps and $\tau_s$ in eight steps. The response curves shown are averaged over time-slides covering the light-travel time of 10\,ms between H1 and L1 using 41 ``tiny" time-slides. (A movie of this injection analysis has been published online.\footnote{van Putten, M.H.P.M., 2020, https://zenodo.org/record/4390382; ibid, 2021, https://zenodo.org/record/4601077.})

In this process, an indicator function $\chi=\chi(t_s,\tau_s,f_s,f_0)$ represents a normalized (H1,L1)-correlation strength over strips along family of curves Eq. \ref{EQN_c} (Paper I, supplementary data). 
This is reduced to $\hat{\chi}(t_s,\tau_s)=\max_{(f_s,f_0)}\chi(t_s,\tau_s,f_s,f_0)$, namely, upon projecting out $(f_s,f_0)$ by maxima over $(f_s,f_0)$ for each $(t_s,\tau_s)$. These projections are necessary by memory limitations, when scanning over a large number of $2\times 10^9$ parameter values per frame of 4096\,s at 4096\,Hz sampling rate.

Response curves vary with $\tau_s$ due to shot-noise in the H1 and L1 detectors noise above a few hundred Hz. Small $\tau_s$ corresponds to fast descent in frequency, putting injections mostly over low frequencies close to $f_0$, where detector noise is less and hence sensitivity is high. Conversely, large $\tau_s$ puts injections close to $f_s$, where shot-noise adversely affects sensitivity. 

\subsection{PFA from event timing based on PDF($t_s$)} 

\begin{figure*}
\centerline{\includegraphics[scale=0.20]{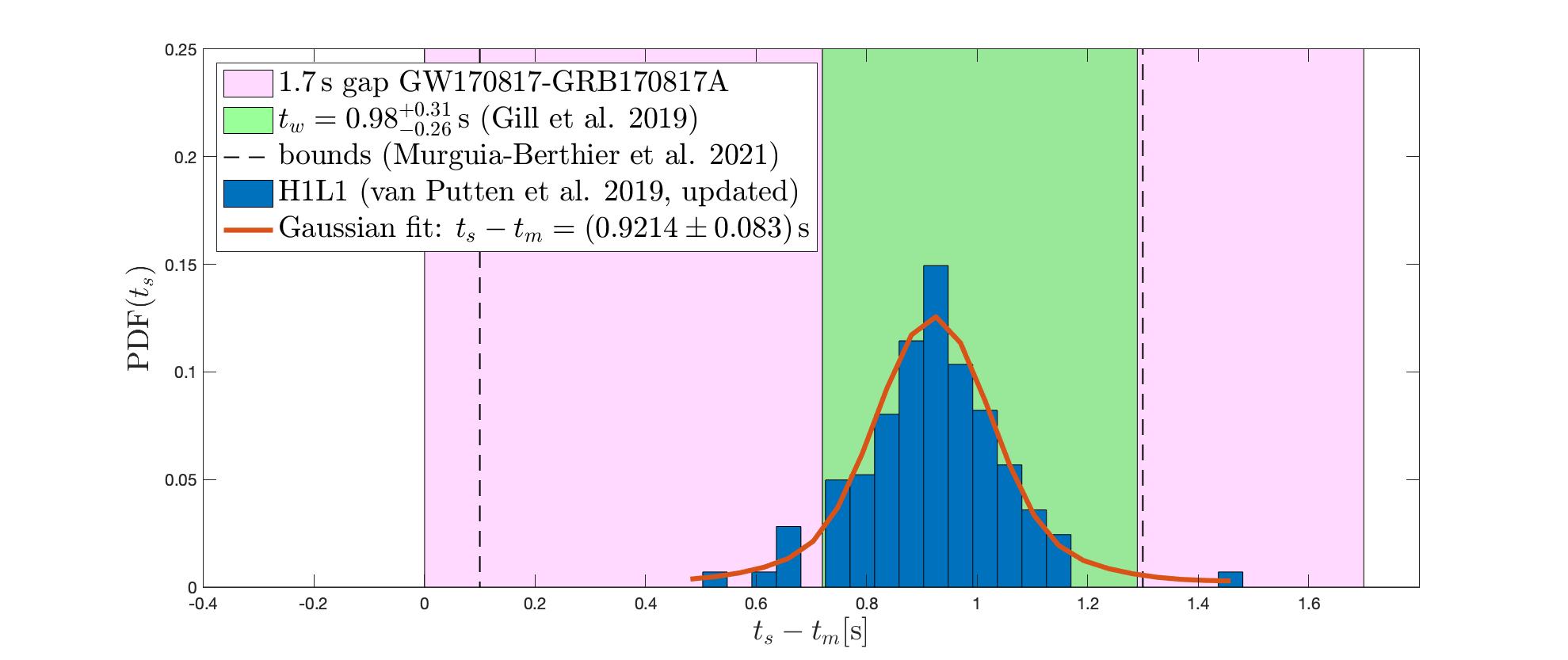}}
\centerline{\includegraphics[scale=0.19]{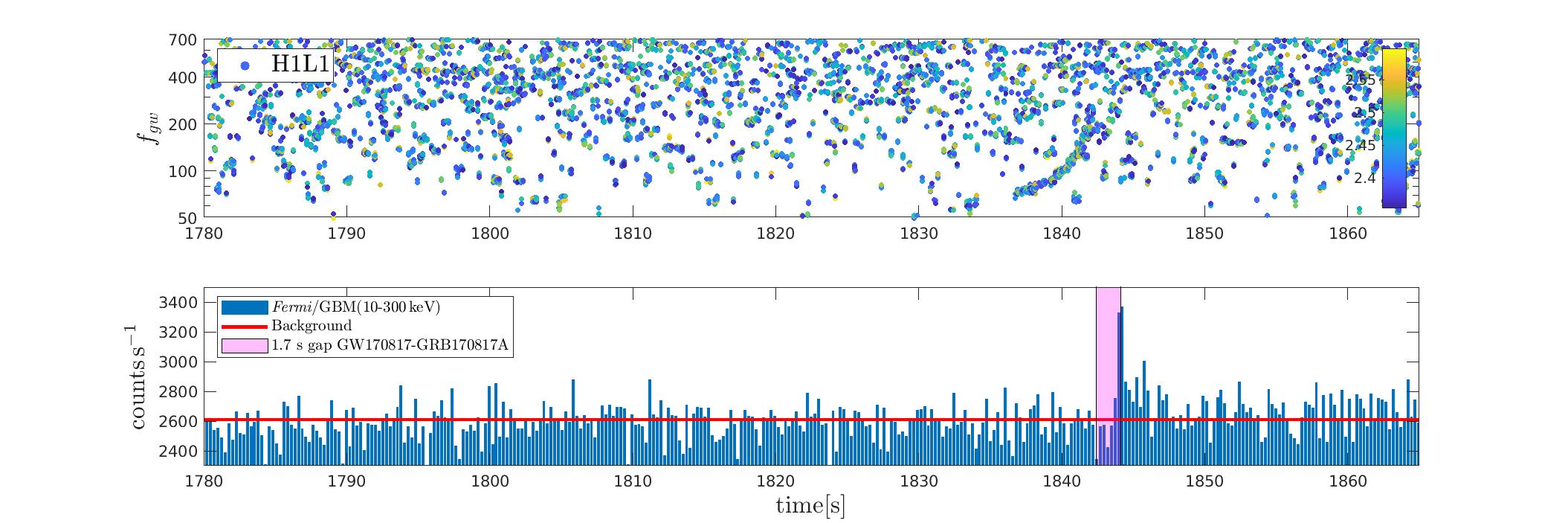}}
\caption{PDF of the start time $t_s$ (blue) of post-merger gravitational radiation to GW170817 following rejuvenation in gravitational collapse to a rotating black hole, setting the initial data to GRB170817A (top-left panel). The PDF shown (bin size 0.05\,s) is superposed to estimates of the lifetime of the progenitor hypermassive neutron star formed in the immediate aftermath of the merger (reviewed in \citep{mur20}). Highlighted is the 1.7\,s gap (magenta) between merger time $t_m=1842.43$\,s and the onset of GRB170817A. Post-merger emission to GW170817 appears as a descending chirp in the (H1,L1)-spectrogram merged by frequency coincidences $\left| f_{\mbox{\tiny H1}}-f_{\mbox{\tiny L1}}\right|<10\,$Hz (lower panels). For the merged spectrogram, parameter estimation (Fig. \ref{figCX2}) derives from time-slides less than the duration $\tau$ of the templates (\ref{EQN_Dt}) against a background generated by larger time-slides.  Post-merger gravitational radiation is identified with the formation of the central engine of GRB170817A, whose duration is constrained by the lifetime of black hole spin (bottom panel). 
} 
\label{fig_tws}
\label{figEE}
\end{figure*}

Significance of a candidate event - satisfying extremal clustering (\S3.3, below) - is estimated by causality in formation of a central engine of GRB170917A (Paper I and II): 
the gap condition (\ref{EQN_C1}) $t_m<t_s<t_m+t_g$, $t_g=1.7\,$s, between GW170817 and GRB170817A, schematically indicated in Fig. \ref{figC1}. 

Including causality in butterfly matched filtering with template duration $\tau$, we consider the slightly more restrictive gap condition, 
\begin{eqnarray}
{C_1}^\prime:~\tau < t_s-t_m < t_g
\label{EQN_ts}
,\end{eqnarray}
in the reduced gap $G^\prime$ of width $t_g-\tau$. Statistical significance is expressed by the PFA: Type I errors in $C_1^\prime$ according to $t_s$ satisfying $t_k(t_s)=1$.

Formally, as per \S1, $t_k(t): [0,T]\rightarrow \{0,1\}$ is a Boolean random variable over our snippet of H1L1-data of duration $T=2048\,$s.  In the present case, one such candidate event (Fig. \ref{figCX2}) marked by the (unique) global maximum at event time $t=t_s$ in an indicator function $\chi(t)$ over $[0,T]$, satisfying extremal clustering over extended foreground $S_1$. Under the null-hypothesis of noise only and no signal, $t_1(t)$ satisfies a uniform prior on $0\le t \le T$. The Type I error, $t_1(t)=1,$ in finding $t_s\,\epsilon\,G$ (or $G^\prime)$ (satisfying $C_1$ (or $C_1^\prime$) by mere chance) is determined by $G$ as a subset of $[0,T]$. Upon partitioning of $[0,T]$ into $n=1204$ intervals of size $\left|G\right|=t_g$, this Type I error carries a PFA given by the ordinary probability $p_1=1/n =t_g/T$. We note that this estimate is conservative compared to the same derived from $G^\prime$, since $\left|G^\prime\right|=t_g-\tau$.

Next, we  generated PDF($t_s$) by application of $\chi$-image analysis over extended foreground: the start-time $t_s$ of a candidate emission feature according to Eq. \ref{EQN_c} in our spectrograms. For descending chirps characteristic for spin-down of a compact object (neutron star or black hole) powering GRB170817A, Eq. \ref{EQN_c} appears minimal in the ``1+3" parameters $t_s$ and (nuisance parameters) $(\tau_s,f_s,f_0$).
Fig. \ref{fig_tws} shows PDF($t_s)$ to be entirely in the $t_g=1.7$\,s gap between GW170817 and GRB170817A.
Under the null-hypothesis - a background of uncorrelated noise in H1 and L1 -- over the original snippet of $T=2048$\,s of H1L1-data, the gap condition $C_1$ is represented by a Boolean random variable $t_k(t)$ with values that are either "true" or "false," that is: $t_k(t)=1$ if $t\,\epsilon\,G$ and $t_k(t)=0$ otherwise. 

Given $G$, $t_s$ in $G$ according to PDF($t_s)$ (Appendix A) carries a PFA $p_1$ equal to Pr$\left(t\,\epsilon\,G\right)$ satisfying (Paper I):
\begin{eqnarray}
p_1=\frac{t_g}{T} = \frac{1.7}{2048}=8.3\times 10^{-4}.
\label{EQN_p1}
\end{eqnarray}
Causality over a background $T$ hereby carries a PFA with two-sided Gaussian-equivalent significance of $3.33\,\sigma$ and false alarm rate (FAR) of about 1 per month, defined by $\mbox{FAR} = \mbox{PFA}/T$ \citep[e.g.,][]{wil18}.

\subsection{Extremal clustering in extended foreground}

Parameter estimation PDF($t_s)$ (also PDF($\tau_s)$) for our candidate feature is derived from merged (H1,L1)-spectrograms over an extended foreground ($S_1$) by application of small time-slides in the interval,
\begin{eqnarray}
I: \left|\Delta t\right|<\tau,
\label{EQN_Dt}
\end{eqnarray}
where $\tau\simeq 0.5$\,s is the intermediate duration of the templates in butterfly matched filtering. We apply this to the original snippet of $T=2048\,$s of H1L1-data in 99 steps symmetric about zero, including a neighborhood of zero covered by 41 tiny time-slides within 10\,ms.

\begin{figure*}
\centerline{\includegraphics[scale=0.183]{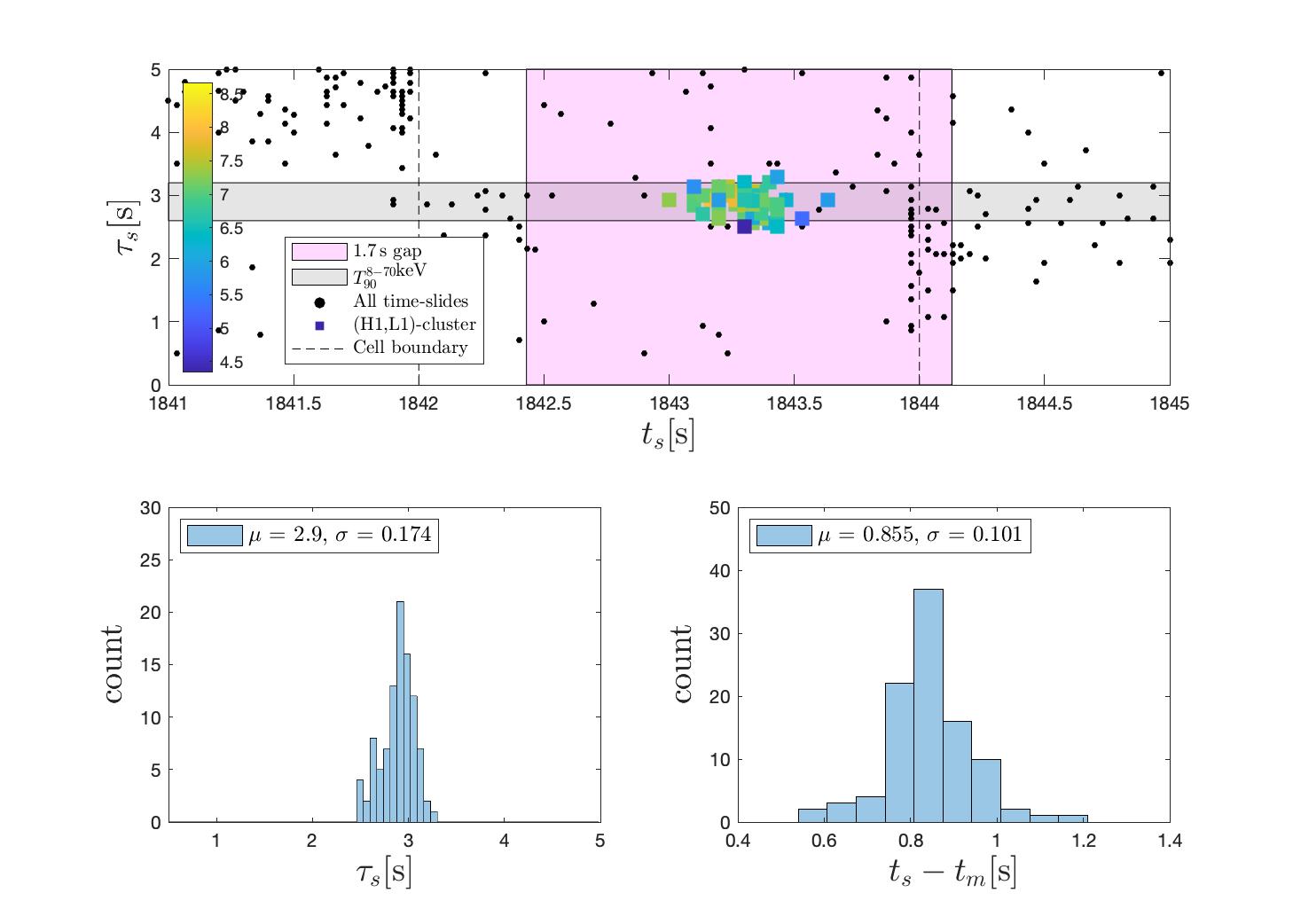}\includegraphics[scale=0.115]{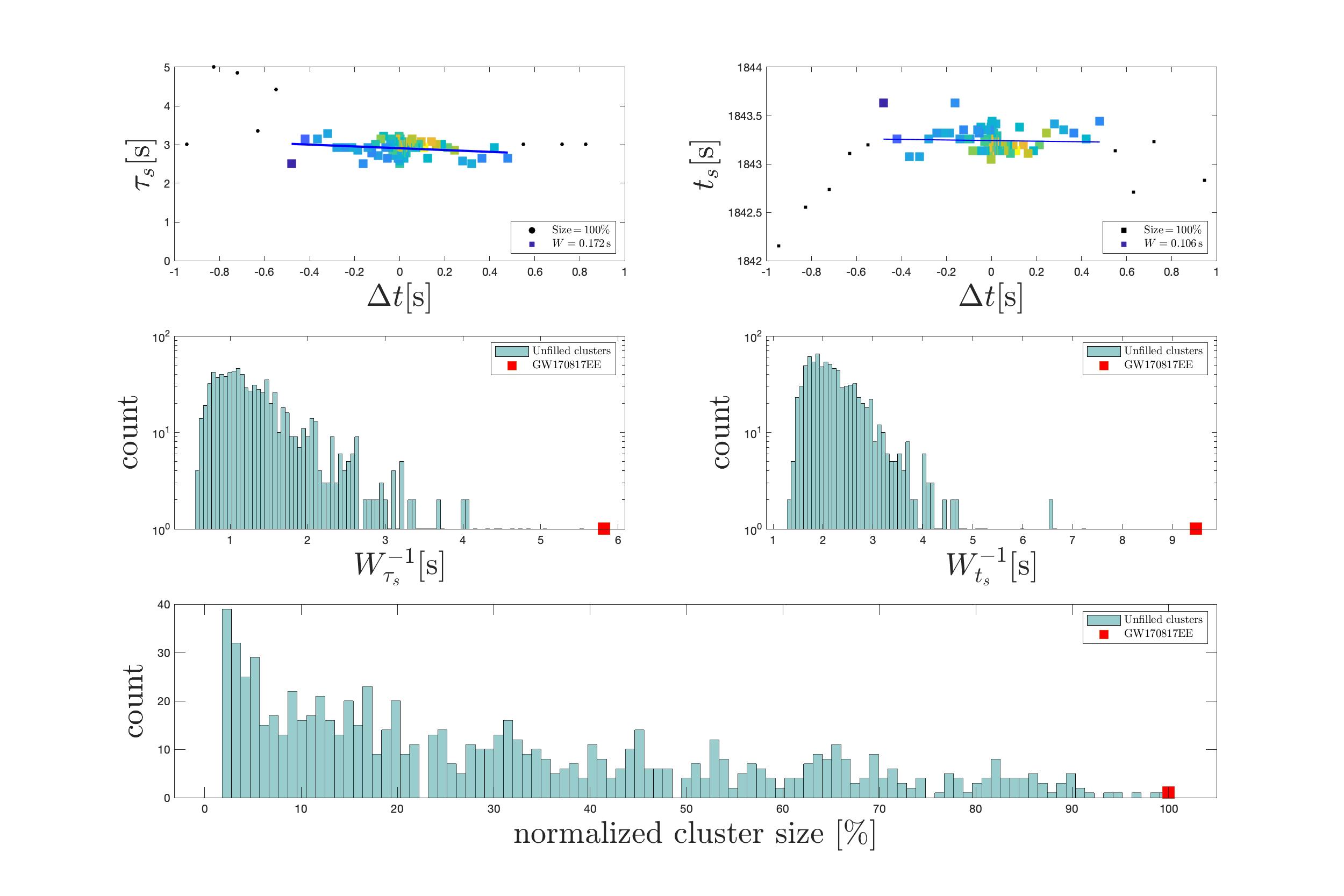}}
\centerline{\mbox{~~~~~~~~}\includegraphics[scale=0.195]{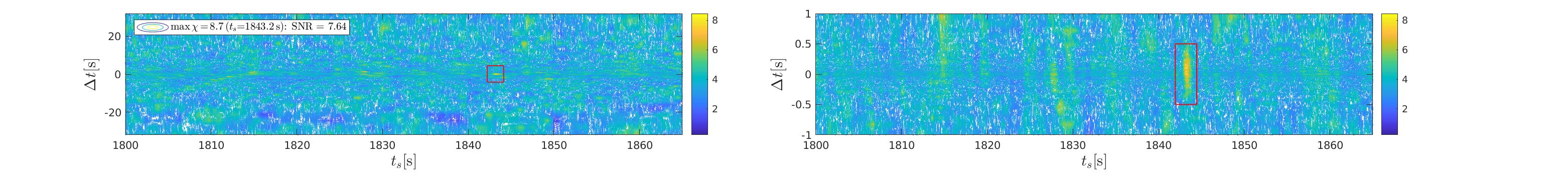}}
\caption{
Joint PDF$(t_s,\tau_s)$ in Eq. \ref{EQN_c} from merged (H1,L1)-spectrograms by $\chi$-image analysis following time-slide analysis (top-left panels). Clustering is seen in the $(t_s,\tau_s)$-plane about a global maximum $\hat{\chi}^*=8.76$ in the extended foreground over small time-slides Eq. \ref{EQN_Dt}. 
(The peak in the cluster in extended foreground ($S_1$) is slightly larger than what is seen in conventional foreground ($S_0$).) This cluster is extremal in the sense of Eq. \ref{EQN_eta}. Tracing produces PDF's of $\tau_s$ and the delay time $t_s-t_m$, where $t_m$ is the merger time of GW170817. There is notable consistency in $\tau_s$ with the duration $T_{90}^{8-70\mbox{keV}}=2.9\pm 0.3\,$s of GRB170817A \citep{poz18}, expected from the timescale of black hole spin-down in interaction with surrounding high-density matter following gravitational collapse of the hyper-massive neutron star in the immediate aftermath of this merger. 
Clustering in $\tau_s$ and $t_s$ as a function of time-slide $\Delta t$ in the post-merger emission to GW170817 extracted from (H1,L1)-spectrograms merged by frequency coincidences $\left| f_{\mbox{\tiny H1}}-f_{\mbox{\tiny L1}}\right|<10\,$Hz (top-right panels). Contour plots of $\hat{\chi}(t_s,\Delta t)$ (maximal over $\tau_s,f_s,f_0$) over all time-slides $\Delta t \,\epsilon\, [-32,32]$\,s ($S_{0-2}$) in the bottom panels. The right panel zooms in on $\Delta t \,\epsilon\, [-1,1]$\,s, highlighting clustering within $\Delta t\,\epsilon\, [-\tau,\tau]$ set by the intermediate time scale $\tau\simeq 0.5$\,s of our template bank. Crucially, this cluster is extremal in the sense of Eq. \ref{EQN_eta} in extended foreground ($S_1$). 
}
\label{figCX2}
\label{figMC}
\end{figure*}

Following Eq. \ref{EQN_Dt}, the parameter estimations of candidate features give rise to "clustering" about peaks $\hat{\chi}^*=\hat{\chi}(t_s^*,\tau_s^*)$ shown in Fig. \ref{figCX2}, where the remaining two parameters $(f_s,f_0)$ in our $\chi$-image analysis have been projected out by taking maxima. Such results by $\chi$-image analysis of spectrograms for each $\Delta t$. In the extended foreground thus produced, clustering in data points -- the output of $\chi$-image analysis of merged (H1,L1)-spectrograms -- is identified in the box,
\begin{eqnarray}
B: \left|\tau_s-\tau_s^*\right|<\tau,~~ \left|t_s-t_s^*\right|<\tau. 
\label{EQN_box}
\end{eqnarray}

For the background, the ensembles of peaks $\hat{\chi}^*$ are identified over a partitioning of a data in cells of duration $w$. Each cell contains a total of $Q=161$ data points (one for each time-slide). According to Eqs. \ref{EQN_Dt}-\ref{EQN_box}, up to 99 may be clustered. 

Accordingly, on the extended foreground ($S_1$) defined by domain (\ref{EQN_Dt}), our two-level pipeline of (H1,L1)-spectrograms followed by $\chi$-image analysis defines a map, whose image may be partially or wholly contained in the box (\ref{EQN_box}). In case of the latter, the normalized cluster size: 
\begin{eqnarray}
\eta = 100\%\times \frac{\left|B\right|}{\left|I\right|}
\label{EQN_eta}
,\end{eqnarray}
defined by the ratio of counts in $B$ and counts in $I$. The limiting case $\eta=100\%$ will be referred to as extremal, shown in Fig. \ref{figCX2}.
(A movie of this time-slide analysis has been published online.\footnote{van Putten, M.H.P.M., 2019, https://zenodo.org/record/3544143.})

Absent a signal, $t_s$ and $\tau_s$ are uncorrelated. A uniform probability distribution of the event time $t_s$ over the data at hand effectively prevents clustering. In response to a signal, $t_s$ and $\tau_s$ tend to be clustered about some central value associated with a goodness of fit by Eq. \ref{EQN_c}. 
Such clustering is representative for correlations between H1 and L1 imparted by a gravitational-wave passing by (Fig. \ref{figCX2}). Clustering in spectrograms hereby increases with signal strength (\cite{van17} and tends to persistent over small time-slides Eq. \ref{EQN_Dt}. 
By this property, the condition of extremal clustering 
provides a robust guard against false positives.

With essentially no free parameters, the extremal condition over extended foreground ($S_1$) provides a robust alternative to conventional thresholding in amplitude-based criteria over foreground. A common implementation of the latter is a pre-defined threshold of signal-to-noise ratio (S/N) of a continuously valued statistic.

Figure \ref{figCX2} with extremal clustering in $(t_s,\tau_s)$ over Eq. \ref{EQN_Dt} confirms identification based on a global maximum $\hat{\chi}^*=\hat{\chi}(t_s^*,\tau_s^*)$ at $\Delta t = 0$ reported previously in Paper I, signaling Extended Emission post-merger in Fig. \ref{figEE}. This cluster is extremal following Eq. \ref{EQN_eta}. Importantly, it is unique over the entire H1L1-data snippet of $T=2048$\,s following a partition over 1024 cells of 2\,s cells. We hereby identify:
\begin{eqnarray}
t_s=(1843.29\pm0.1)\,\mbox{s},
\label{EQN_ts0} 
\end{eqnarray}
satisfying $C_1^\prime$ - a more precise statement of $C_1$. In Eq. \ref{EQN_ts0} and in what follows, results are quoted by central value and STD of the PDF. 

Estimating PDF($t_s)$ by Eq. \ref{EQN_c} serves to produce a statistic $t_s$ in which $(\tau_s,f_s,f_0)$ are nuisance parameters. Evaluated over aforementioned moving data-segments of $\Delta T$ obtains well-defined event timing $t_s$ at the maximum of $\hat{\chi}(t_s)$, similar though not identical to deriving a stable estimate of the duration of light curves of long GRBs in the face of large temporal variability by means of matched filtering \citep{van09}.

Given the uniform prior in timing (in the absence of astrophysical context), Eq. \ref{EQN_ts} is satisfied with associated probability of false alarm Eq. \ref{EQN_p1}. Tracing the cluster of Fig \ref{figCX2} generates PDF's of $t_s$ and $\tau_s$, demonstrating:
\begin{eqnarray}
t_s-t_m =\left( 0.86\pm0.10\right)\mbox{s},~\tau_s =\left( 2.91\pm0.17 \right)\mbox{s},
\label{EQN_dts1}
\end{eqnarray} 
where the first follows from Eq. \ref{EQN_ts0}.

Robustness of maximal $\hat{\chi}$ conditional on extremal clustering in Fig. \ref{figCX2} pans out in uniqueness already in a minimally extended foreground of 7 time-slides extending over $T^\prime \simeq  100\,T$ in analyzing 50 frames of H1-L1 data, blindly selected to avoid biases in data-quality. This background analysis increases significance (\ref{EQN_p1}) in timing by causality by a factor of $T^\prime/T$, that is, improving (\ref{EQN_p1}) by: 
\begin{eqnarray}
p_1^\prime =\frac{t_g}{T^\prime}=\frac{1.7}{204800}=8.3\times 10^{-6}
\label{EQN_p1b}
,\end{eqnarray}
with a two-sided Gaussian-equivalent significance of $4.46\,\sigma$ and FAR of about 1 per 782 years. This degree of robustness motivates a further analysis of timing based on individual analysis of H1 and L1 data with no regards to context. Starting point for is generating PDF's in parameter estimation circumventing the use of time-slides, to be applied to each of the H1 and L1 data-channels.

\section{Analysis of individual H1 and L1 spectrograms}
 
\begin{figure*}
\begin{center}
\centerline{
\includegraphics[scale=0.142]{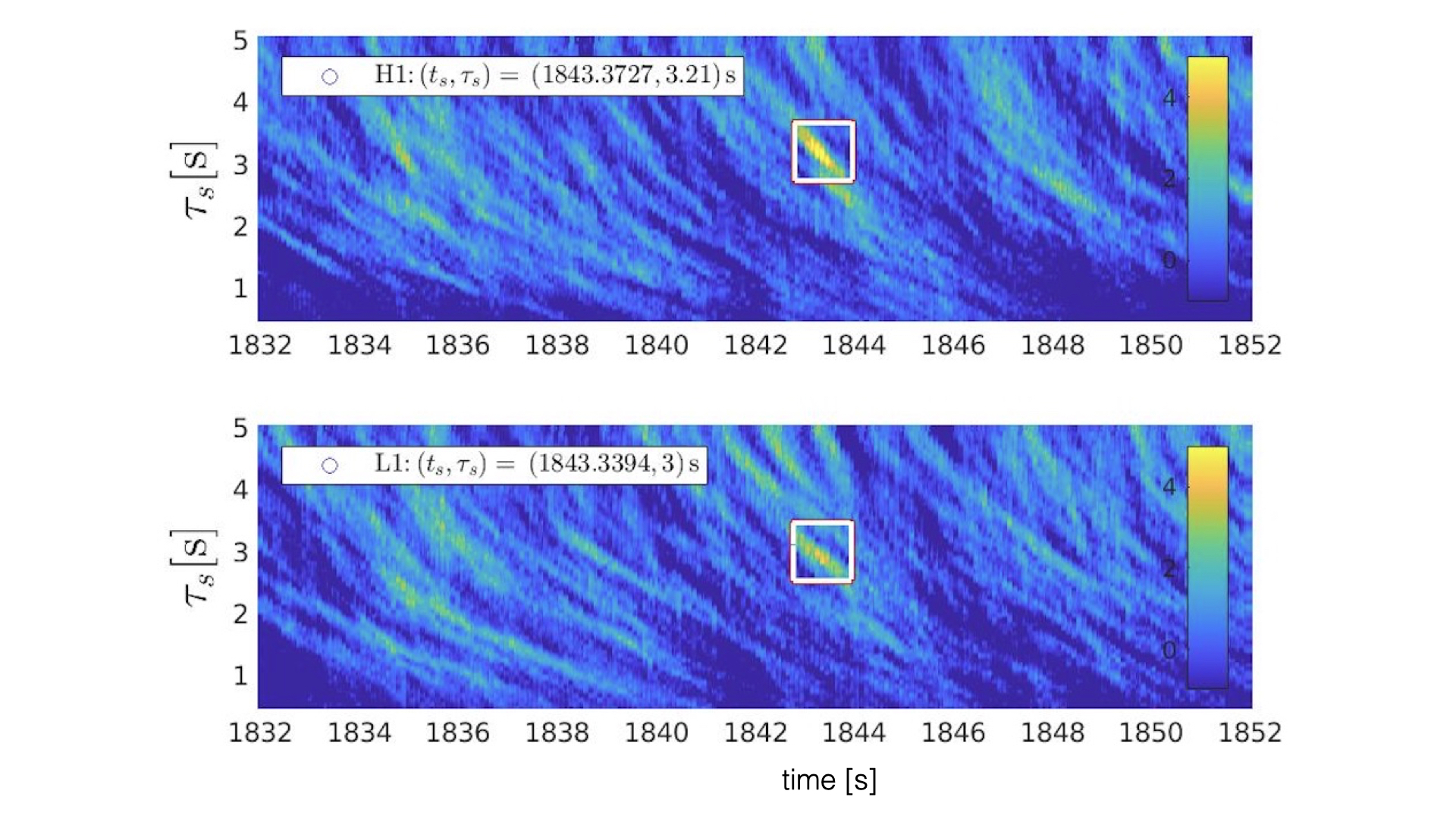} 
\includegraphics[scale=0.142]{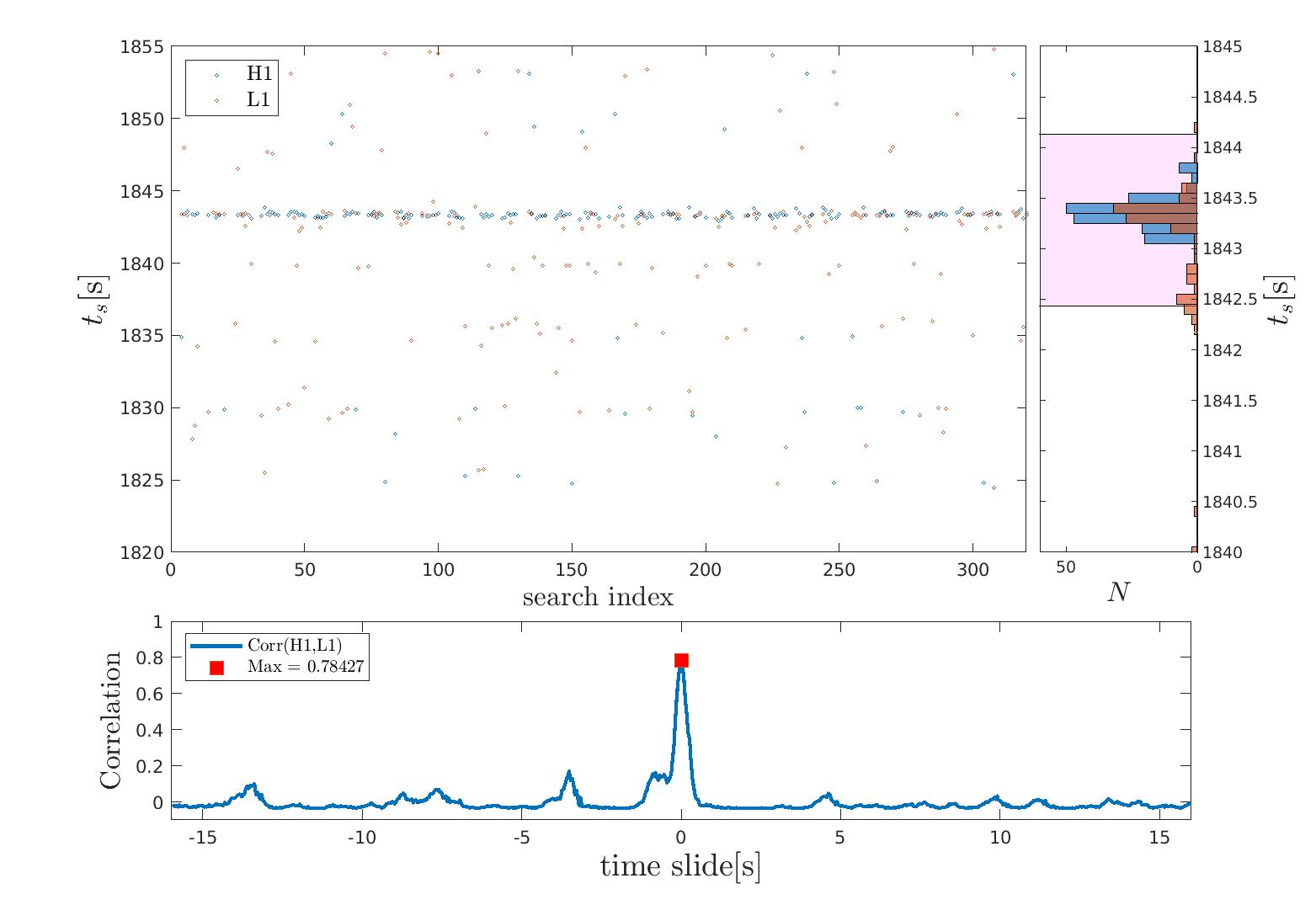}}
\centerline{\hskip0.3in\includegraphics[scale=0.25]{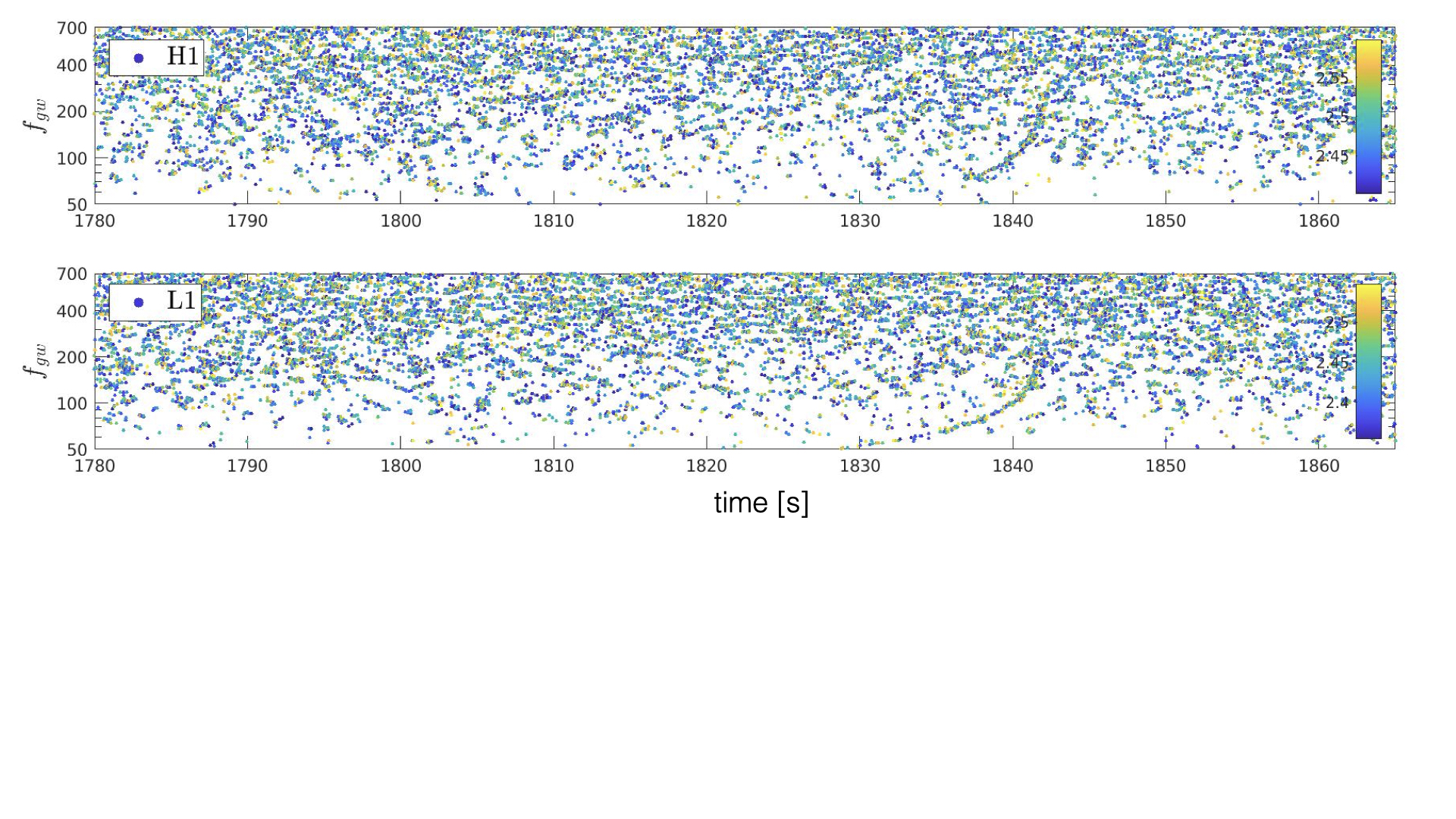}}
\vskip-1.5in
\end{center}
\caption{
Sample of ($t_s^*,\tau_s^*)$ identified in $\chi$-image analysis of individual H1- and L1-spectrograms (top-left panels), representing two elements from a total of $nm=320$ H1- and L1-spectrograms (lower panels). 
Peak values indicated (white rectangles) have $\hat{\chi}=7.0$ (H1) and $\hat{\chi}=5.3$ (L1).
Over the snippet of $T=2048\,$s of H1L1-data, time differences $\left[t_s\right]= t_{s,{\mbox{\tiny H1}}}^*-t_{s,{\mbox{\tiny L1}}}^*$ and $\left[ \tau_s \right]= \tau_{s,{\mbox{\tiny H1}}}^*-\tau_{s,{\mbox{\tiny L1}}}^*$ identified by local maxima of $\hat{\chi}\left(t_s,\tau_s\right)$ over the 64 segments of 32\,s and 320 trials of H1- and L1-spectrograms are, as expected, uncorrelated, evidenced by a correlation coefficient $\rho_{\left[ t_s\right],\left[ \tau_s\right]} = 1.29\times 10^{-4}$. 
(Right top panels.) PDF$_1(t_s)$ and PDF$_2(t_s)$ of H1 and, respectively, L1 are centered around $t_s-t_m=\left(0.92\pm 0.09\right)\,\mbox{s}$ and $\tau_s = \left(3.00 \pm 0.09\right)\,\mbox{s}$. Binning over $\delta t_{res} = 0.025\,$s, cross-correlation of the two PDFs (over each segment covering H1L1-data over $T=2048$\,s) shows a global maximum in segment 58 containing GW170817 consistent with zero time difference within $\delta t_{res} = 0.025$\,s.}
\label{figTSPDF}
\end{figure*}

Independent of context, statistical significance in timing can be evaluated according to consistency in $t_s$ obtained from the mutually independent H1 and L1 observations:
\begin{eqnarray}
{C_2}:~\left[ t_s \right] \equiv  t_{s,{\mbox{\tiny H1}}} -  t_{s,{\mbox{\tiny L1}}} = 0,
\label{EQN_dts4}
\label{EQN_C2}
\end{eqnarray}
within the finite precision of our observation. 

To evaluate $C_2$, we appeal to PDF$_1(t_s)$ of H1 and PDF$_2(t_2)$ of L1 and their cross-correlation function. The cross-correlation of the two will have a maximum, whose proximity to zero will be determined by the associated STD. As before, given a total duration of observation, $T$, this STD normalized to $T$ provides us with a second PFA. This PFA from $[t_s]$ is independent of the PFA($t_s$) in the mean of H1 and L1 (\S3).

Absent gravitational waves passing by, H1 and L1 output are uncorrelated. Given this null-hypothesis, mean and difference of timing $\left( t_s, \tau_s\right)$ are independent. A PDF$\left(t_{s,{\mbox{\tiny H1}}}, t_{s,{\mbox{\tiny L1}}}, \tau_{s,{\mbox{\tiny H1}}},\tau_{s,{\mbox{\tiny L1}}}\right)$ hereby factorizes as $p_A\left(t_s,\tau_s \right) p_B\left(\left[ t_s \right], \left[ \tau_s \right] \right)$. 
Integration over differences reduces it to $p_A\left(t_s,\tau_s\right)$; and to (\ref{EQN_p1}) when further integration over $\tau_s$. 
On the other hand, integrating over $\left(t_s,\tau_s\right)$ reduces the same to $p_B\left(\left[ t_s \right], \left[\tau_s\right]\right)$ independently of the context, that is, with no regards to the multimessenger data GW170817-GRB170817.

The finite dispersion in the cross-correlation of PDF$_1(t_s)$ and PDF$_2(t_s)$ is expected to be quite small, albeit greater than the light travel-time between H1 and L1. In butterfly matched filtering over templates of intermediate duration $\tau$, temporal resolution in a single cross-correlation between data and template is governed by $\tau$. However, resolution is enhanced when averaging over a number of trials. To this end, we generated PDF's by $nm$ trials in a search extending over $n$ seeds of our template banks in butterfly matched filtering, while output is branched into $m$ spectrograms defined by a stride $m$. Here, template seeds refer to the master templates prior to time-slicing over $\tau$ 
and conversion to time-symmetric templates \citep{van14a}. As in the analysis of the previous section, these calculations are realized by modern heterogeneous computing (Appendix D).

\begin{table*}[]
\caption{Results on event timing in post-merger emission 
Eq. \ref{EQN_2} to GW170817 identified with rejuvenation Eqs. \ref{EQN_ratio} in Eq. \ref{EQN_MS}. Type I errors derive independently from the gap condition, $C_1$ (PFA $p_1$), and mutual consistency, $C_2,$ (PFA $p_2$) in H1L1-data over a duration, $T$. {Central values and uncertainties derive from} mean 
and standard deviation of PDF($t_s$) and PDF($\tau_s$). 
{Uncertainties defined by the width of the respective PDFs reflect amplitude information.}}  
\begin{center}
{\tiny 
\begin{tabular}{|c|c|c|c|c|c|c|c|c|c|c}
\hline
\hline
& & & & & & &\\
{\bf  H1} & {\bf L1} & {\bf H1}  & {\bf L1} & {\bf H1,L1}  & {\bf H1,L1}  &     {\bf merged (H1,L1)}  & {\bf merged (H1,L1)}\\ & & & & & & &\\
\hline
$t_s-t_m$[s] & $t_s-t_m$[s] & $\tau_s$[s] & $\tau_s$[s] & $t_s-t_m$[s] & $\tau_s$[s] &    $t_s-t_m$[s] & $\tau_s$[s]\\ 
$0.9130\pm0.1366$ & $0.9234\pm0.1122$ & $3.1\pm0.1$ & $2.9\pm0.2$ & $0.92\pm 0.09$ & $3.00\pm0.09$ &   $0.86\pm 0.10$ & $2.91\pm0.17$\\
\hline
{\bf $T$[s]} & {\bf $p_1$} & {\bf FAR$^{-1}_1$} & {\bf $p_2$} & {\bf FAR$^{-1}_2$}      & {\bf $p_1 p_2$}  &  {\bf FAR$^{-1}$} & \\
\hline 
2048 & ${1.7}/{2048}$     & 1 month & ${0.1}/{2048}$ & 1.4 yr                          &  $4.1\times 10^{-8}$  & 1.6\,kyr & \\  
204800 & ${1.7}/{204800}$ &  782 yr& ${0.1}/{2048}$    & 1.4 yr                     &   $4.1\times 10^{-10}$  & $>160\,$kyr & \\  
\hline
\hline
\end{tabular}
}
\end{center}
\label{T1}
\end{table*}

We generate PDFs with a total of $nm=320$ elements from $n=10$ different template seeds and a stride $m=32$. 
From a single template seed, the output of butterfly filtering comprises $L={\cal O}\left(10^8\right)$ lines
of output of the form:
\begin{eqnarray}
\left(i,\rho_{1},\rho_{2},f_{11},f_{11},f_{21},f_{22}\right)_k
\label{EQN_L}
,\end{eqnarray}
where $t_i=i/4096$\,s is the sample time, $\rho_{i}$ denotes the correlation between data and specific template for detector H1 $(i=1)$ and L1 $(i=2)$ and $f_{ij}$ refers to the initial $(j=1)$ and final $(j=2)$ frequency of the template. For a stride $m$, this output branches into $m$ spectrograms. Image analysis of each hereby produces a PDF of the parameters at hand, thus broadening our search by a factor $nm$ in each of H1 and L1.

Inherent to our approach is a bank of templates that is dense rather than orthogonal, and approximate over an intermediate times scale \citep{van14a,van17}. A sample $i$ in (\ref{EQN_L}) is hereby typically covered by a multiplicity of different templates. The average time-stepping: 
\begin{eqnarray}
\Delta t_{ik} = t_{i_{k+1}} - t_{i_{k}},
\end{eqnarray}
in $t_i$ is correspondingly smaller than the sample time $1/4096$ s of data; $\Delta t_{ik}$ is typically zero or $1/4096$\,s with STD of about 40\,$\mu$ s. A stride $m=32$ hereby covers about 7 samples. The associated STD in the mean of frequencies is similarly small, i.e., ${\cal O}\left(10^{-2}\right)$ Hz. Offsets $m_0=1,2,\cdots,m$ hereby provide exceptionally high-resolution time-stepping in our matched filtering analysis.

Figure \ref{figTSPDF} shows the results of parameter estimation in the $(t_s,\tau_s)$-plane of H1 and L1 by $\chi$-image analysis applied to the H1- and L1-spectrograms, generated over a large number of trials. Noticeably, our candidate signal appears stronger in H1 than in L1, indicated by $\hat{\chi}(t_s^*)$ and the peak in the H1- and L1-histograms over $t_s$. Furthermore, the results over $T=2048\,$s of H1L1-data for $\left[t_s\right]= t_{s,{\mbox{\tiny H1}}}^*-t_{s,{\mbox{\tiny L1}}}^*$ and $\left[ \tau_s \right]= \tau_{s,{\mbox{\tiny H1}}}^*-\tau_{s,{\mbox{\tiny L1}}}^*$ at local maxima of $\hat{\chi}\left(t_s,\tau_s\right)$ satisfy a correlation coefficient of $\rho_{\left[ t_s\right],\left[ \tau_s\right]} = 1.29\times 10^{-4}$, as expected from otherwise independent operation of H1 and L1. Consequently, we have the factorization $p_B=p_2 \left( \left[ t_s \right]\right)  \times p_3 \left( \left[ t_s \right]\right)$, 
facilitating integrating out either $\left[ \tau_s \right]$ or $\left[ t_s \right]$. 

Figure \ref{figTSPDF} shows PDF$_1(t_s)$ and PDF$_2(t_s)$ thus extracted from H1 and L1 with their cross-correlation obtained by time-slide analysis applied to this pair. 
A key finding is that these PDF's are maximally correlated at a time-slide consistent with zero within the time-slide resolution $\delta t_{res} = 0.025\,$s. 
This maximum is global (maximal over all such maxima identified in the 64 segments of 32 s covering our analysis of the H1L1 data of
duration $T=2048\,$s. It thereby carries a PFA:
\begin{eqnarray}
p_2=\frac{4\delta t_{res}}{T} = 4.88 \times 10^{-5}
\label{EQN_p2}
,\end{eqnarray}
with equivalent two-sided Gaussian-equivalent significance of $4.06\,\sigma$ and FAR of $\sim1$ per 1.4 year.

In deriving Eq. \ref{EQN_p2} from $\left| \left[ t_s \right]\right| < \delta t_{res}$ at the peak of cross-correlating PDF$\left( t_{s,{\mbox{\tiny H1}}}\right)$ and PDF$\left( t_{s,{\mbox{\tiny L1}}}\right)$, a factor of 2 derives from absolute value of time-differences $\left[ t_s \right] =\left|t_{\mbox{\tiny H1}} - t_{\mbox{\tiny L1}}\right|$ 
in H1 and L1; another factor of 2 derives from the top-hat distribution in $\left[  t_s \right] $ based on uniform priors in $t_{\mbox{\tiny H1}}$ and $t_{\mbox{\tiny L1}}$. Similar to $p_1$, consistency in the PDF's of $t_s$ is hereby formally cast under the null-hypothesis by the Boolean random variable $Y\,\epsilon\{0,1\}$ with $Y=1$ if $\left| \left[ t _s \right]  \right| < \delta t_{res}$, $Y=0$ otherwise. 

The mean, now from the pair of PDFs of $t_s$ from the independent H1- and L1-analyses, shows a delay time in gravitational collapse
\begin{eqnarray}
t_s-t_m=\left(0.92\pm 0.09\right)\,\mbox{s},
\label{EQN_dts2}
\end{eqnarray}
consistent with Eq. \ref{EQN_dts1} based on the mean $t_s$ in merged (H1,L1)-spectrograms. 

In astrophysical context, Eq. \ref{EQN_dts2} is consistent with the lifetime $0.98^{+0.31}_{-0.26}$\,s of the progenitor hypermassive neutron star derived from jet propagation times and the mass of blue ejecta \citep{gil19} and its survival time, based on neutrino cooling and internal dissipation of rotational energy \citep[e.g.][]{ben21}.
For $\tau_s$, the PDFs associated with the cluster (\ref{EQN_dts2}) show $\tau_s=\left(3.1\pm 0.1\right)\,$s (H1) and $\tau_s=\left(2.9\pm 0.2\right)\,$s (L1) with mean:
\begin{eqnarray}
\tau_s = \left(3.00 \pm 0.09\right)\,\mbox{s},
\label{EQN_taus2}
\end{eqnarray} 
consistent with Eq. \ref{EQN_dts1} in the previous analysis of merged (H1,L1)-spectrograms. Identified with the lifetime of black hole spin, it appears to constrain the duration $T_{90}^{8-70\mbox{keV}}$ of GRB170817A \citep{poz18}. Thus, Eqs. \ref{EQN_dts2}-\ref{EQN_taus2} both appear to possess natural astrophysical context in the electromagnetic spectrum.

\section{Discussion}

The complex merger sequence Eq. \ref{EQN_MS} is probed by multimessenger calorimetry and event timing (Fig. \ref{figC1}). 
Following calibration by signal injection experiments (Fig. \ref{figRes}), a post-merger descending chirp characteristic for spin-down of a compact remnant is analyzed for output ${\cal E}_{GW}$ and statistical significance by event timing subject to two consistency conditions, $C_{1,2}$, on time-of-onset $t_s$ seen in merged and individual H1- and L1-spectrograms (\S3-4). 
Figure \ref{figEC1} shows a detailed summary.

${\cal E}_{GW}$ in Eq. \ref{EQN_2} suffices to break the degeneracy between a hyper-massive neutron star and a black hole by exceeding limits in $E_J$ of the first. Yet, it is amply accounted for by $E_J$ following (delayed) gravitational collapse to a rotating black hole in the aftermath of GW170817 (\S2).
Statistical significance is expressed by a PFA factorized over two statistically independent PFAs Eq. \ref{EQN_p1} in \S3 and Eq. \ref{EQN_p2} in \S4, derived from consistency conditions on effectively the mean ($C_1$) and, respectively, a difference ($C_2$) in start-time, $t_s$.

\subsection{Insensitivity to numerical parameters}

Event timing analysis is rather insensitive to numerical parameters given that certain discrete conditions are met in PDF($t_s)$ when evaluating $C_1$ (Appendix A) and in considering the difference $\left[t_s\right]$ in $C_2$ as follows. PFA $p_1$ from $C_1$ (more conservative than PFA from $C_1^\prime$) is determined by $G$ relative to the observational window $\left[0,T\right]$, given that PDF($t_s)$ is well within $G$ (Fig. \ref{fig_tws}, Appendix A). 
$p_1$ is hereby insensitive to trial factors that might otherwise be encountered when inferred from a statistic in continuous stochastic variables such as the S/N. It likewise does not involve Occam factors, common to likelihood analysis of the same in Bayesian approaches. The PFA $p_2$ from $C_2$ is derived from cross-correlating PDF($t_s$) based on two individual H1 and L1 analyses. Sampled by $nm=320$ trials, they appear to be close to the noise limit of the data. Moreover, any systematic uncertainty in PDF($t_s$) cancels out in the difference $[t_s]$ (\S4).

An additional safeguard derives from extremal clustering (100\% in Eq, \ref{EQN_eta}). It implies the width of PDF$(t_s)$ to be much smaller than the gap size, $t_g$, of GW170817-GRB170817A (Fig. \ref{fig_tws} in \ref{EQN_C0}). 
Over extended foreground, it derives as a goodness of fit over $(t_s,\tau_s,f_s,f_0)$, representing an essentially minimal number of ``1+3" parameters for descending chirps. (By count this is the same in a canonical CBC search, that is, taken over two masses and orbital ellipticity for merger times $0\le t_m\le T$.) There is no fine-tuning in PDF$(t_s)$ derived from Eq. \ref{EQN_c} with maxima $\hat{\chi}=\hat{\chi}(t_s)$ upon projecting out the nuisance parameters $(\tau_s,f_s,f_0)$.

\begin{figure*}
\begin{center}
\centerline{\includegraphics[scale=0.22]{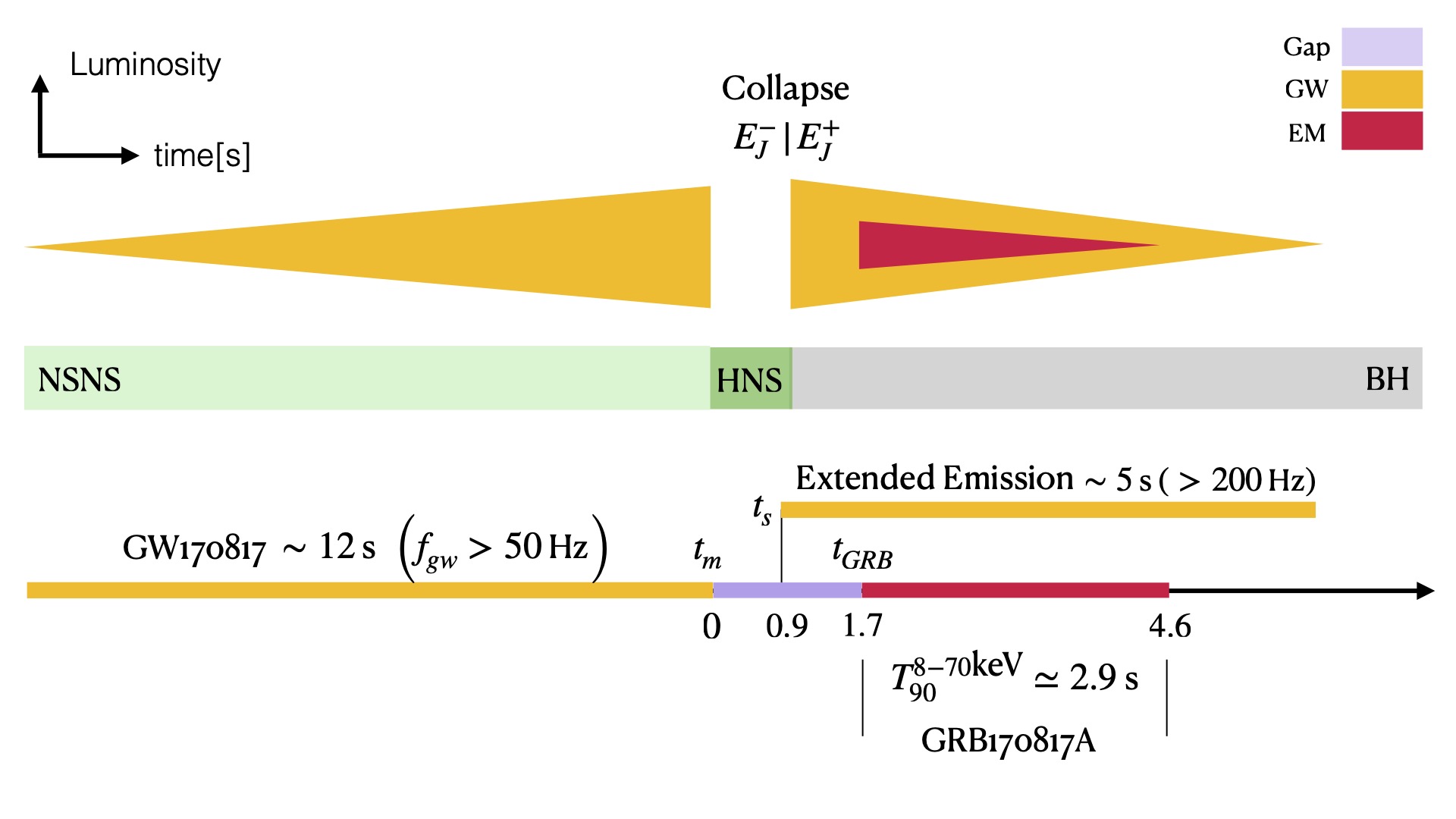}}
\vskip-0.4in
\end{center}
\caption{
Chronicle (\ref{EQN_MS}) showing detailed event timing in Fig. \ref{figC1}. Highlighted is ${\cal E}_{GW}$ by rejuvenation in gravitational collapse of the hyper-massive neutron star formed in the immediate aftermath of the merger at $\lesssim t_s$ in the 1.7\,s gap $G$ between GW170817-GRB170817A, in a Kerr black hole with spin-energy $E_J^+$ exceeding $E_J^-$ in its progenitor according to (\ref{EQN_ratio}). $E_J^+$ amply accounts for ${\cal E}_{GW}=\left(3.5\pm1\right)\%M_\odot c^2$, breaking the degeneracy between neutron stars and black holes.}
\label{figEC1}
\end{figure*}

Our focus on Eq. \ref{EQN_MS} gives a concrete example of multimessenger calorimetry and event timing analysis, summarized in the chronicle Fig. \ref{figEC1}. 
The joint PFA $p_1 p_2 = 4.1\times 10^{-8}$ from (discrete) event timing is noticeably on par with the PFA $8.91\times 10^{-7}$ of GW170817-GRB170817A, merging the two $p$-values of (continuous stochastic variables) in temporal and spatial agreement between H1, L1, and the {\em Fermi} GBM reported by LIGO-Virgo \citep[][]{abb17c}. 
A further third PFA derives from consistency in ${\tau_s}$ from the individual H1 and L1 observations (Table 1). In this regard, our total PFA by $t_s$ is conservative.

\begin{table*}[]
\caption{EM-GW timing consistency in the multimessenger afterglow emission to GW170817 comprising the kilonova AT2017gfo, GRB170817A and a descending branch in GW-emission.}
\begin{center}
{\tiny 
\begin{tabular}{|c|c|c|c|c|c|c|c|c|c|c}
\hline
\hline 
& & & \\
& {\sc Lifetime HNS, start-time descending branch} & {\sc Duration GRB170817A, time-scale of descent} & {\sc References}\\ 
& & & \\
\hline
{\bf EM} & $t_w=\left(0.98\pm 0.3\right)$\,s & $T_{90}^{(8-70){\rm keV}}=\left(2.9\pm 0.3\right)$\,s  & \cite{gil19,poz18}\\ 
{\bf GW} & $t_s = \left(0.92\pm0.083\right)$\,s & $\tau_s=\left(3.0\pm 0.09\right)$\,s & this work \\
\hline
\hline
\end{tabular}
}
\end{center}
\label{T2}
\end{table*}

\subsection{Advances over previous results}

We present two independent analysis of merged and individual H1 and L1 spectrograms, results of which may be compared for consistency and combined to enhance confidence levels on the basis of high-resolution PDFs (\S3-4). To be precise, in the absence of a gravitational-wave signal, data from H1 and L1 are statistically independent (the working hypothesis in multi-detector GW analysis).
Under the null-hypothesis, merged and individually, H1 and L1 satisfy uniform priors on event timing. We consider the mean (\S3) and difference (\S4) in event timing expressed in a statistic $t_s$, representing two linearly independent combinations of the H1 and L1 time-series equivalent to a rotation over $\pi/4$ in the plane. By unitary of rotations, statistical independence of the H1 and L1 data is preserved in the two statistics $t_s$ of \S3-4. Table 1 lists the measurement results.

A comparison with previous results derived from foreground ($S_0$, time-slide zero) is also opportune. Our PDF$(t_s)$ produces $t_s\simeq 0.92$\,s (Table 1), revising $t_s\simeq 0.67$\,s (Paper II). 

This revised estimate is illustrative for a limitation of foreground in representing a single sample in PDF($t_s)$ derived from extended. In the face of finite scatter seen in the plot of $t_s$ versus time-slide $\Delta t$ (Fig. 4, top right panel), foreground estimates incur a statistical uncertainty on the order of the width of PDF($t_s)$. Additionally, foreground incurs a systematic error given by the finite difference in signal arrival time between the two detectors.

The PDF($t_s)$s from merged and individual H1 an L1 analysis permit a direct comparison between the two for their implied PFAs $p_1$ and $p_2$ (Table 1). The inferred PFAs are qualitatively distinct, however, being contextual (by the gap $t_g=1.7\,$s between GW170817-GRB170817) and, respectively, acontextual. Strictly speaking, a direct numerical comparison of the two is not opportune. 

Nevertheless, our $p_1>p_2$ may appear paradoxical in light of a visibly higher contrast of the descending branch in merged rather than individual spectrograms. Indeed, the width in PDF($t_s$) is relatively smaller, approximately by a factor of $\sqrt2$ (Table 1), as expected.
Increasing the time of observation by $\times 100$, $p_1^\prime < p_2$ (Eq. (\ref{EQN_p1b}) with confidence level $4.46\sigma$. 
In fact, including amplitude information, $p_1$ decreases below $p_2$ to $2.5\times 10^{-5}$ ($4.2\sigma$, supplementary data to Paper I). 
The same applied to $p_1^\prime$ provides a combined confidence level of $5.16\sigma$, but this avenue is not pursued further here.

Discrete event timing producing PFAs differently from amplitude-based PFA (Appendix A). Fig. 1 shows the candidate signal to be sufficiently strong for PDF($t_s$) to be well within the gap of GW170817-GRB170817A, satisfying extremal clustering (Fig. 4). It hereby satisfies causality – a Boolean valued conclusion. True, in the case at hand, fixes PFA$_1$ at $p_1=t_g/T$, where $T$ is the time of observation. A further increase in signal strength, narrowing PDF($t_s$), will not change this value.

For uniformity of presentation, PFAs are here given by event timing only, sufficient to derive a significantly improved joint PFA, factored over the independent PFA$_1$ and PFA$_2$, to validate the descending branch.

\subsection{Black hole spin-down}

We identified our estimate of $\tau_s\simeq3\,$s in Eqs. \ref{EQN_dts1}, \ref{EQN_taus2} with the Kelvin-Helmholtz time-scale of black hole spin-down against high-density matter through a torus magnetosphere \citep{van03,van15}:
\begin{eqnarray}
\tau_H=\frac{E_J}{\left|\dot{E}_J\right|} \simeq 3\,\mbox{s}\left(\frac{\sigma}{0.05}\right)^{-1}\left(\frac{z}{6}\right)^4\left(\frac{M}{3M_\odot}\right),
\label{EQN_tauH}
\end{eqnarray}
where $\sigma=M_T/M$ in the right hand-side refers to the mass-ratio of torus-to-black hole with radius $zR_g$, $R_g=GM/c^2$, specialized to the present GRB170817A. This timescale Eq. \ref{EQN_tauH} is derived from the first law of thermodynamics. 

For a black hole with mass, $M$, angular momentum, $J_H$, and angular velocity, $\Omega_H$, interacting with a torus magnetosphere rotating at the angular velocity $\Omega_T$ of the inner face of a surrounding torus, $\tau_H$ in (\ref{EQN_tauH}) largely represents the dissipation, $D_H$ of $E_J$, on the event horizon, leaving a net luminosity 
$L_H=-\dot{M}\ll D_H$ \citep{van03,van15}: 
\begin{eqnarray}
L_H = -\Omega_H\dot{J}_H-D_H = L_j+L_T 
\label{EQN_LT}
,\end{eqnarray}
in the limit of small Reynolds stresses in suspended accretion, where $L_j\ll L_T$ refers to the luminosity $L_j$ in a baryon-poor jet (BPJ) along an open magnetic flux tube subtended over a finite polar angle $\theta_H\ll\pi/2$ on the event horizon of the black hole and a luminosity $L_T\simeq -\Omega_T\dot{J}$ in multimessenger radiation from the surrounding inner disk or torus. Here, the inequality $L_j\ll L_T\simeq L_{GW}$ is seen in discrepant energies in electromagnetic Eq. \ref{EQN_EM1} and, respectively, gravitational radiation Eq. \ref{EQN_2}.

GRB170817A derived from a BPJ is included in Eq. \ref{EQN_EM1}. While GRB170817A is negligible in the total energy budget $E_J^+$ (\S2), it provides crucial timing information $t_{GRB}$ and $T_{90}^{(8-70){\rm keV}}$. It can be attributed to $L_j$ derived from a Faraday-induced potential \citep{van00,van03},
\begin{eqnarray}
U=\omega J_p
,\end{eqnarray}
of charged particles with angular momentum $J_p=eA_\varphi$ along surfaces of constant flux $A_\varphi = \Phi/2\pi$ subject to frame-dragging by a black hole in its lowest energy state. While $L_j$ for $\theta_H\ll\pi/2$ is relatively small, it may have observational consequences for GRBs and UHECRs alike and especially so when intermittent \citep{van09,sha15,van15b,got21}. 
For a discussion on high-energy emission from black holes that are not in the lowest energy state, we refer to \cite{rue22}.

\section{Conclusions}

From analysis of H1 and L1 observations in merged and, new in this work, individual detector spectrograms producing statistically independent PFAs $p_1$ and, respectively, $p_2$, {from their respective high-resolution PDF($t_s)$s} of an extended emission feature, a number of findings emerge:
\begin{enumerate} 
\item A black hole central engine to GRB17017A based on ${\cal E}_{GW}\simeq 3.5\%M_\odot c^2$ 
in a descending chirp 
exceeding the limits of the HNS in the immediate aftermath of GW170817 (Eqs. \ref{EQN_2}, \ref{EQN_c}, \ref{EQN_2b}), attributed to rejuvenation of $E_J$ in gravitational collapse (\S2);
\item Statistical significance with combined PFA of $4\times 10^{-8}$ (equivalent $Z$-score 5.48) factored over two independent PFAs in Eqs. \ref{EQN_p1} and \ref{EQN_p2} derived from event timing subject to $C_{1,2}$ (Table 1);
\item A delay time $t_s-t_m\lesssim1$\,s in post-merger gravitational collapse in agreement with independent estimates of the lifetime of the hypermassive neutron star (Fig. \ref{fig_tws}, Table 2), ruling out a BH-NS merger progenitor \citep{cou19};
\item A timescale of descent $\tau_s\simeq 3$\,s identified with the lifetime of black hole spin in agreement with the duration $T_{90}^{8-70\mbox{keV}}=2.9\pm 0.3\,$s of GRB170817A \citep{poz18}. We refer in particular to Figs. \ref{figCX2}, \ref{figEC1}, and Table 2;
\item The gravitational-wave energy emitted in the combined ascending-descending chirp of GW170817 is about 2\% of the total mass-energy of the system.
\end{enumerate}

\section{Outlook}

Planned LVK {observational runs O4-5} offer potentially important new observational opportunities to probe merger sequences involving neutron stars and central engines of energetic core-collapse supernovae in the Local Universe. EM-GW observations involving GW-calorimery and event timing offer some new tools to identify their nature {\citep{cut02}, particularly when breaking the degeneracy between (super- or hyper-) massive neutron stars and black holes.}

Since black holes have no memory of their progenitor except for total mass and angular momentum, such appears notably opportune for type Ib/c supernovae as the parent population of normal long GRBs \citep{van19c} and superluminous supernovae \citep{don15}. 
While rare, failed GRB-supernovae nevertheless may be luminous in gravitational radiation and more frequent than double neutron star mergers.

For the planned LVK runs O4-5, core-collapse supernovae may be probed in blind or optically triggered searches in the Local Universe over distances comparable to GW170817. If detected, gravitational-wave emission is likely to identify their enigmatic central engine, significantly complementing our understanding of core-collapse events currently limited to SN1987A. In the more distant future, searches for broadband gravitational-wave radiation may put rigorous and model-independent bounds on (non-axisymmetric) mass-motion around supermassive black holes such as SgrA* by the planned Laser Interferometric Space Antenna \citep[LISA;][]{van19c}.
\begin{acknowledgements}
The authors gratefully acknowledge a detailed reading and constructive comments from the anonymous reviewer and M.A. Abchouyeh, which greatly contributed to clarity of presentation.
The first author gratefully acknowledges stimulating discussions with Gerard 't Hooft over a Nico van Kampen Colloquium on GW170817 at ITP, University of Utrecht (2019). LIGO O2 data are from the LIGO Open Science Center of the LIGO Laboratory and LIGO Scientific Collaboration (LSC), M. Vallisneri et al., 2014, Proc. 10th LISA Symp., University of Florida, Gainesville (May 18-23), arXiv:1410.4839, funded by the U.S. National Science Foundation. The original GW170817 2048\,s data set is 10.7935/K5B8566F of the LIGO Laboratory and the LSC. Virgo is funded by the French Centre National de Recherche Scientifique (CNRS), the Italian Instituto Nazionale della Fisica Nucleare (INFN) and the Dutch Nikhef, with contributions by Polish and Hungarian institutes. Additional support is acknowledged from MEXT, JSPS Leading-edge Research Infrastructure Program, JSPS Grant-in-Aid for Specially Promoted Research 26000005, MEXT Grant-in-Aid for Scientific Research on Innovative Areas 24103005, JSPS Core-to-Core Program, Advanced Research Networks, and the joint research program of the Institute for Cosmic Ray Research. Computations have been performed on a 
dedicated platform by synaptic parallel computing for dynamical load balancing. This research is supported, in part, by NRF of Korea Nos. 2018044640 and 2021K1A3A1A16096820. MdV acknowledges support from PRIN-MIUR 2017 No. 20179ZF5KS.
The data underlying this article were accessed from the LIGO Open Science Center https://www.gw-openscience.org/about/, specifically the 2048\,s data set 10.7935/K5B8566F containing the merger GW170817. Derived data generated in this research will be shared on reasonable request to the corresponding author.
\end{acknowledgements}

%
%

\begin{appendix} 

\section{Discrete event timing versus S/N}

GW170817-GRB170817A provides a multimessenger context by the gap $G$ of $t_g=1.7\,$s in between these two events with merger time $t_m$ and, respectively, the time-of-onset $t_{GRB}$ (Fig. \ref{figC1}).

For a candidate post-merger emission feature associated with the central engine of GRB170817A, a PFA derives from event timing as an alternative to conventional signal-to-noise (SNR) analysis under the null-hypothesis of stationary detector noise {\em and} the uniform prior of astrophysical event timing ($H0$).

To illustrate this, consider an indicator function $\chi(t)$ marking the event time $t_e$ of a candidate feature by its global maximum over a finite duration of observation $T$. Since $\chi(t_e)$ is continuous stochastic observable, $t_e$ is well-defined and unique. Thus, $t_e$ tends to be uniformly distributed over $[0,T]$, namely:\  
\begin{eqnarray}
{\rm Pr}_0 = p(t_e|H0)=\frac{1}{n}
\label{EQN_Pr0}
,\end{eqnarray}
with event time now discrete over $n=T/t_g$ bins. 
The gap condition (\ref{EQN_C1}) hereby carries a PFA equal $p_1=t_g/T$. 

In contrast, gravitational-wave emission from the putative central engine of GRB170817A ($H1$) changes our expectation to ${\rm Pr}=p(t_e|H1)$ with a bias to satisfying Eq. \ref{EQN_C1}, provided $\chi(t)$ is suitably devised to measure the start-time $t_e=t_s$ of this emission. As may be expected, $p(t_e|H1)>p(t_e|H0)$ for $t_e\,\epsilon\,G$ depends on the S/N.

\begin{figure}
\centerline{\includegraphics[scale=0.18]{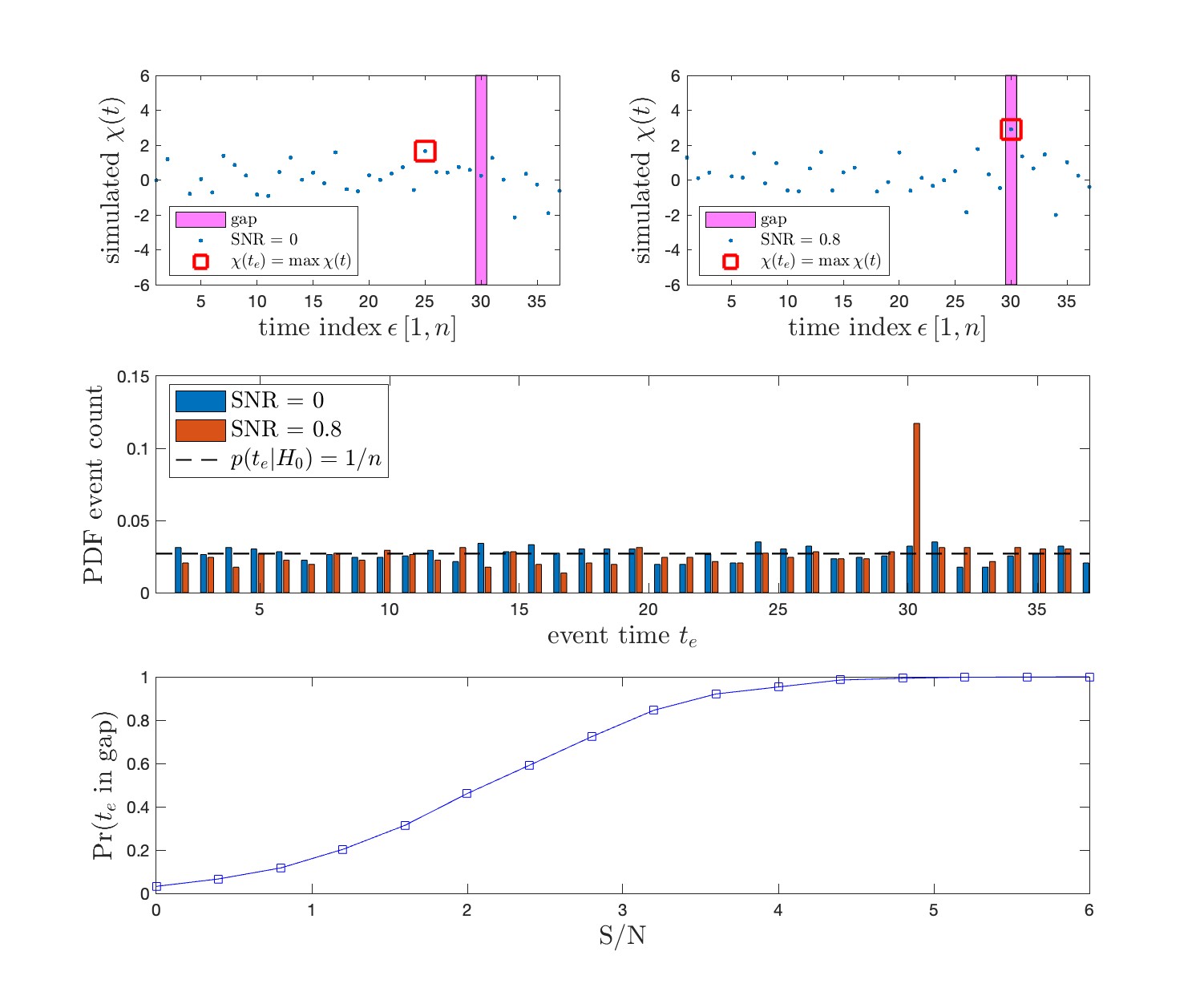}}
\caption{
MC simulation on discrete event timing over $n=37$ cells associated with a gap condition at cell 30, shown in magenta (top and middle panels). The event time, $t_e$, of a single run is the location of the maximum of an indicator function, $ \chi(t),$ over all cells. 
A signal from a source in the gap ($H1$) increases the probability, $p(t_e|H1),$ of detection at this location, above the PFA $p(t_e|H0)=1/n$ of stationary background noise ($H0$). The probability $p(t_e|H1)$ increases monotonically with S/N, here derived from 1024 runs (lower panel).
}
\label{figA0}
\end{figure}

Figure \ref{figA0} shows an elementary Monte Carlo simulation of the probability Pr$\left(t_e\,\epsilon\,G\right)$ of satisfying the gap condition Eq. \ref{EQN_C1} as a function of S/N, based on event times, $t_e$, defined by maxima of $\chi(t)$ given by the sum of normally distributed noise over $n=37$ bins plus a signal at bin 30.\footnote{van Putten, M.H.P.M., 2022, https://zenodo.org/record/7185535}
For $n=37$, this PFA equals the chance of winning a single bet in European roulette conform the probability theory of Blaise Pascal.

A PFA from Eq. \ref{EQN_C1} rather than based on the S/N has the advantage of being independent of trial factors, though the result is a priori limited by $G$ (relative to $T$) and the implementation requires PDF($t_e$) to be sufficiently narrow relative to $G$, that is:\ 
\begin{eqnarray}
\sigma_{t_e}\ll t_g \ll T.
\label{EQN_C0}
\end{eqnarray}

Once (\ref{EQN_C0}) is satisfied, $p_1={\rm Pr}_0$ is independent of the width of PDF($t_s$), which might otherwise depend on trial factors. In this sense, (\ref{EQN_C0}) represents a discrete timing condition. As an ordinary probability, furthermore, $p_1$ readily combines with other PFAs.

\section{Whitening H1L1-data}

\begin{figure*}
\includegraphics[scale=0.47]{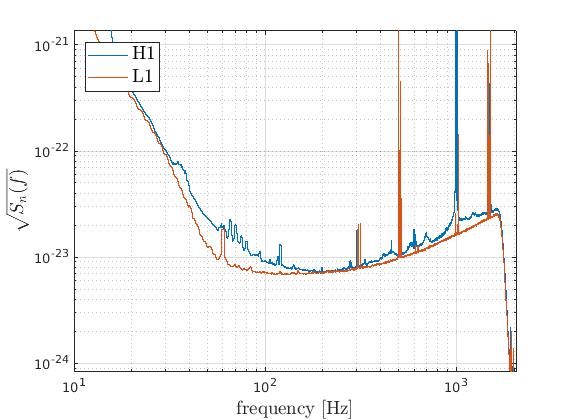}
\includegraphics[scale=0.47]{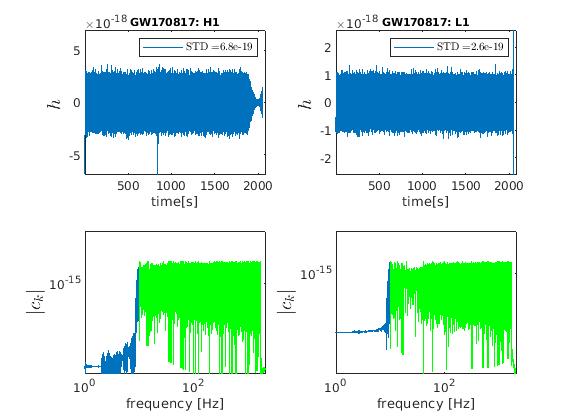}
\caption{
(H1L1-detector spectral energy density $S_n(f)$ for the $T=2048$\,s data snippet containing GW170817 following a band pass filter 10-1700\,Hz (left panel). Violin modes appear prominently associated with the suspension of optics. Frequencies up to about 1700 Hz can be used in injection experiments. H1 and L1 detector noise is very similar though not identical during the GW170817 event.
Whitening of H1 and L1 strain data by normalisation (\ref{EQN_Ck}) produces flat spectra with lines suppressed (right panels). 
}
\label{figA1}
\end{figure*}

As a pre-processing step applied to LIGO strain data, we applied a band pass filter of 10-1700\,Hz followed by whitening to suppress numerous lines, mostly violin modes which appear prominently in the spectrum of the detectors (Fig. \ref{figA1}) \citep{abb20}. 

Figure \ref{figA1} shows the computation by the standard Welch method. Here, the frequency resolution is slightly less than in Fig. 1 of Paper II by a different partitioning in the time domain. By the time-frequency uncertainty relation, lines hereby vary in width yet with the same total energy by Parseval’s Theorem. We note, however, that constant frequencies are suppressed in our butterfly matched filtering (Appendix C).

To this end, strain data are normalized in Fourier domain by amplitude following a partitioning of the spectrum over intervals of $B$\,[Hz] (supplementary data, Paper I).
Specific to the $T=2048$\,s snippet of H1L1-data covering GW170817 sampled at a $n_s=4096$\,Hz, we have $N=2^{12}[T/{\rm s}]=2^{23}$ Fourier coefficients $c_k$ $\left(k=1,2,\cdots N\right)$. Its Fourier spectrum is partitioned up to the Nyquist frequency $\frac{1}{2}n_s$ by intervals $B_s$: $(s-1)M < k \le s\times M$ of size $B$[Hz], each comprising $M=\left(B/n_s\right)N=BT$
coefficients; to exemplify this, $M=4096$ for $B=2\,$Hz. Normalization to an essentially flat spectrum is obtained by:
\begin{eqnarray}
C_k=\frac{c_k}{A_s}~~\left(k\,\epsilon\,B_s\right)
\label{EQN_Ck}
,\end{eqnarray}
over all intervals $s=1,2,\cdots,\frac{1}{2}n_s/B$, where $A_s={M}^{-1}\sum_{k\epsilon B_s}\left|c_k\right|$
is the mean of the absolute values of $c_k$ in $B_s$ with $\left|B_s\right|=B$. Crucially, (\ref{EQN_Ck}) preserves phase in each of the Fourier coefficients. Effectively the same whitening obtains upon setting $A_s$ equal to the standard deviation of the $c_k$ in $B_s$. 

This procedure suppresses violin modes in the LIGO detectors, provided $B$ is about 1-10\,Hz - $B$ must exceed the maximal width of the violin modes yet be relatively modest to effectively suppress the same. The result is essentially flat spectra (Fig. \ref{figA1}), while preserving signals of interest such as the relatively long-duration merger signal GW170817, evidenced in Fig. \ref{fig_tws}. Whitening (\ref{EQN_Ck}) is effective also in rendering GW170817 to be directly audible, for instance, using the MatLab function sound.

In the present application to the H1L1-data covering GW170817, we empirically verified in previous work (Paper I)
that all results remain essentially unchanged for $B=2-16\,$Hz.

\section{Butterfly matched filtering}

\begin{figure}
\hskip-0.8in
\includegraphics[scale=0.185]{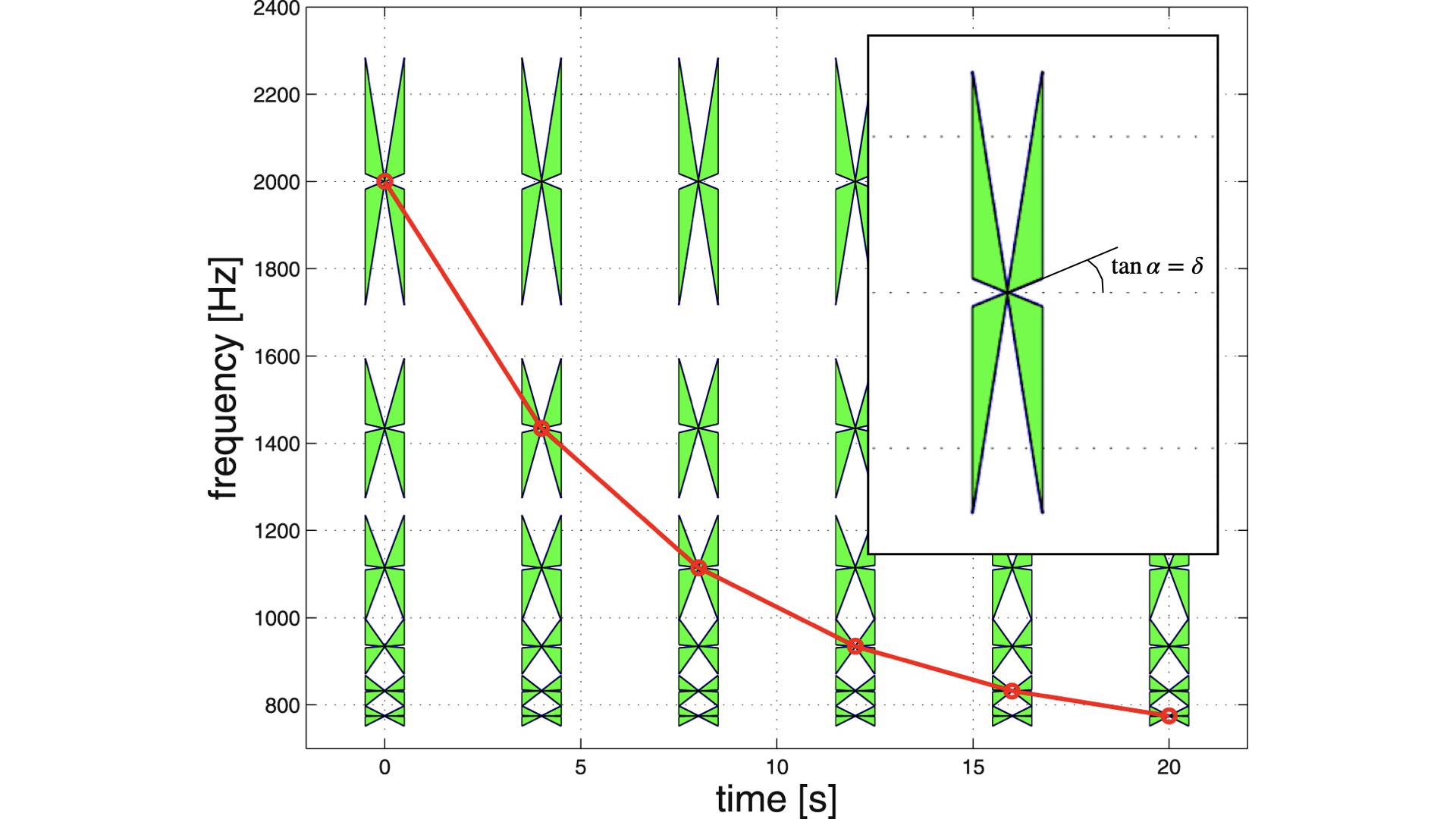}
\centerline{\includegraphics[scale=8.00]{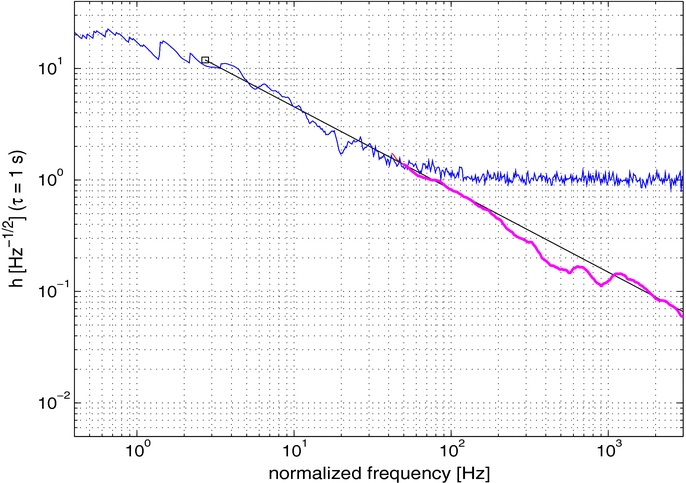}\mbox{~~~~~~~}}
\caption{
Time-frequency domain illustrating filtering signals with frequencies slowly wandering in time, that appear as trajectories with finite time-derivative (top panel). Schematically, these trajectories passing through ``butterflies" (green) defined by a finite slew rate (time rate-of-change) in frequency $\left|df(t)/dt\right|\ge\delta > 0$ (insert). It is realized by matched filtering over a dense bank of chirp-like time-symmetric templates of intermediate duration. (Reprinted from \cite{van16}.)
 Identification of broadband Kolmogorov spectrum in an ensemble averaged spectrum to the Nyquest frequency of 1\,kHz of 42 bright long GRBs in the BeppoSAX catalog by butterfly filtering over a dense bank of 8.64 million templates, shown as a purple line (bottom panel). This result demonstrates sensitivity to turbulence better than an order of magnitude compared to conventional Fourier analysis (blue). Note:\ reprinted from \cite{van14a}.
}
\label{figA2}
\end{figure}

Butterfly filtering is an essentially linear filter for signals with frequencies slowly wandering in time parameterized by $\delta$ (Fig. \ref{figA2}). This is devised by matched filtering over an effectively dense bank of time-symmetric chirp-like templates of duration $\tau$, intermediate to the period and the total duration of a candidate signal \citep{van17}. 

In the time-frequency diagram, the passing of such signals can be schematically indicated by "butterflies" (Fig. \ref{figA2}), indicated by a finite slew rate (time rate-of-change) in frequency: 
\begin{eqnarray}
\left|\frac{df(t)}{dt}\right|\ge \delta
,\end{eqnarray}
for some choice of $\delta>0$. Constant frequency signals fail to pass through the same, indicating relative suppression. Butterfly matched filtering is hereby distinct from Fourier analysis, that favors passing signals with relatively constant frequency. 

Butterfly filtering was originally developed to extract broadband spectra of light curves of long GRBs of the {\em BeppoSAX} catalog \citep{van14a}. It identifies Kolmogorov spectrum which extends up to the Nyquist frequency of 1024\,Hz of the {\em BeppoSAX} sampling rate of 2048\,Hz (during the first 8 seconds of long bursts), demonstrating a sensitivity one order of magnitude beyond what is attained by conventional Fourier analysis (Fig. \ref{figA2}). 

\begin{figure}
\centerline{\mbox{~~~~}\includegraphics[scale=7.8]{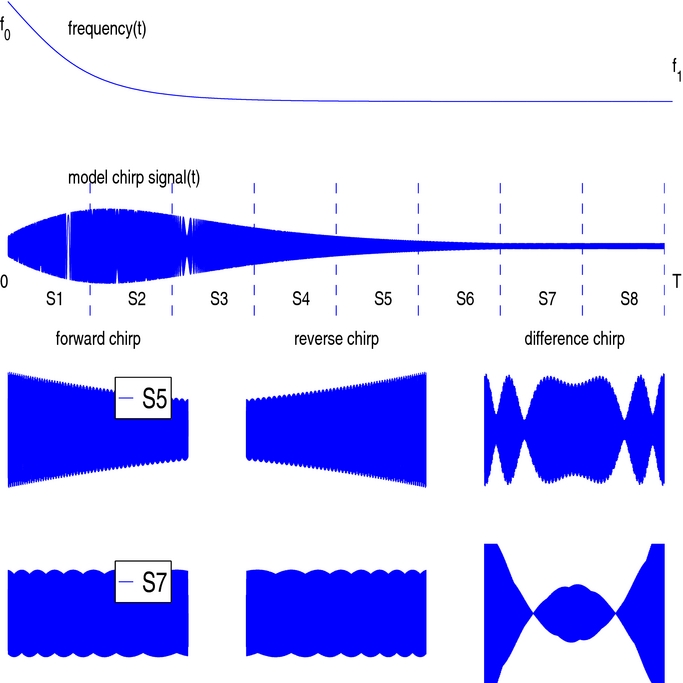}}
\centerline{\includegraphics[scale=0.14]{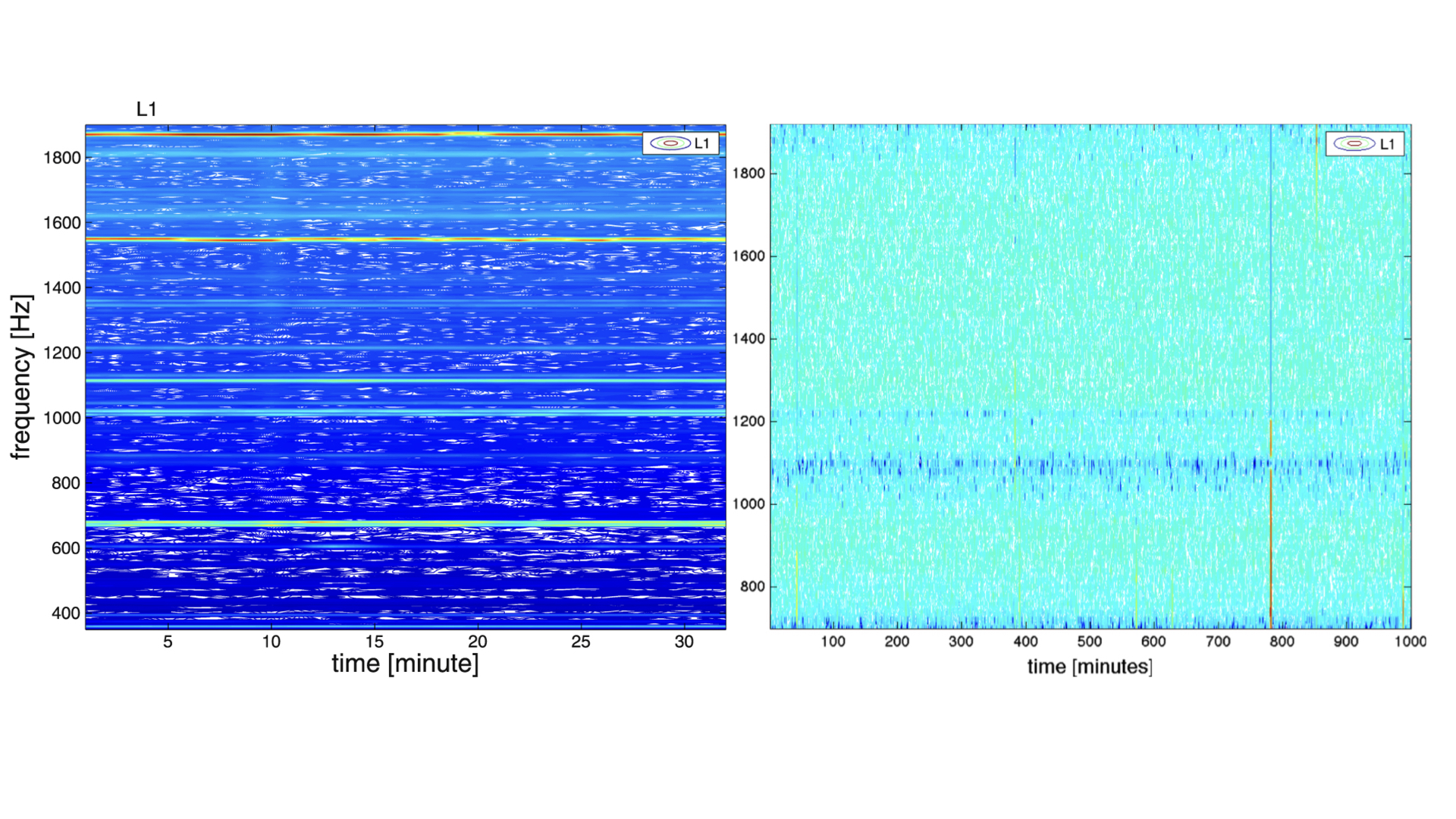}}
\vskip-0.29in
\caption{
 Master template serving as a seed to the construction of a template bank, given by an exponential decay in frequency $f_0\ge f(t) > f_1$ over $0\le t \le T$ (top panel). Time-slicing produces chirps of intermediate duration $P\ll \tau \ll T$, where $P$ denotes the characteristic period of the (candidate) signal of interest. Time-symmetric templates obtain by linear superposition with their time-reverse (right), realizing equal sensitivity for ascending and descending chirp-like signals. (Reprinted from \cite{van14a}.)
Illustration of the suppression of constant frequency input (lower panel) in the form of the application of butterfly filtering to unwhitened LIGO data (left). Lines such as shown in Fig. \ref{figA1} fail to correlate with butterfly templates, leaving a smooth and approximately white background noise (right). Note:\ reprinted from \cite{van16}.
}
\label{figA3}
\end{figure}

The template bank of butterfly matched filtering densely covers a two-parameter range in frequency and time rate-of-change of frequency. These templates are produced by time-slicing of a seed template (Fig. \ref{figA3}) over aforementioned duration $\tau$. Time-symmetric templates obtain by linear combination with their time-reverse, realizing equal sensitivity to signals whose frequency increases or decreases in time. 

Butterfly filtering is ported to LIGO data analysis with no principle change in algorithm. An early demonstration shows the suppression of lines in unwhitened LIGO data, as constant frequency signals fail to correlate with chirp-like templates (Fig. \ref{figA3}). 

Analysis of large data-sets using a dense bank typically containing $2^{18}$ templates requires heterogeneous computing, off-loading matched filtering to graphics processor units (GPUs) with additional procedures to manage limitations of bandwidth between GPU and host CPU given exascale computations involved \citep[][Appendix C]{van17}. 

\section{Heterogeneous Parallel Computing}

\begin{table*}[]
\caption{List of symbols and key words in the two-level pipeline of H1L1 data analysis probing the merger sequence Eq. \ref{EQN_MS} subject to constraints $C_{1,2}$ evaluated by PDF$(t_s)$ produced by butterfly matched filtering and $\chi$-image analysis, augmented by time-slide analysis, implemented by load balanced heterogeneous parallel computing.}  
\begin{center}
{\tiny \begin{tabular}{|c|cc|}
\hline\hline
{\bf  Symbol}.  & {\bf Definition} & {\bf Comment} \\ 
\hline
{\sc GW170817-GRB170817A} & & \S1-2\\
\hline 
$t_m$[s]        & {\rm Merger time GW170817} & Fig. \ref{figC1}\\
$t_g$[s]        & {\rm Gap in GW170817-GRB18017A} & Fig. \ref{figC1}\\
${\cal E}_{EM}$ & {\rm EM output} & Eq. \ref{EQN_EM1}\\
${\cal E}_{GW}$ & {\rm GW output} & Eq. \ref{EQN_2}\\ 
\hline
{\sc Post-merger Emission} & & \S2-4\\ 
\hline
$t_s$[s]    & {\rm Start time} & Eq. \ref{EQN_c}\\ 
$\tau_s$[s] & {\rm Time-scale of decay in the time-frequency domain} & Eq. \ref{EQN_c}\\
$f_s$[s]    & {\rm Initial frequency} & Eq. \ref{EQN_c}\\
$f_0$[s]    & {\rm Late-time frequency} & Eq. \ref{EQN_c}\\ 
\hline
{\sc Butterfly matched filtering} & & \S3-4\\ 
\hline 
$\tau$      & Duration of templates in butterfly matched filtering & Fig. \ref{figCX2} \\
$\rho$      & Matched filtering output & Normalized in post-callback to clFFT\\
$\kappa$    & Threshold of $\rho$ output to host & Applied in post-callback to clFFT \\ 
\hline
{\sc Spectrograms} & & \S3-4\\
\hline 
(H1,L1)-spectrogram & Spectrograms merged by frequency coincidences &  \S3, Figs. 2-3 \\
$\Delta t$          & Time-slide applied to H1-L1 & control parameter\\ 
$S_0$               & Foreground & $\Delta t = 0$ \\
$S_1$               & Extended foreground & $\left|\Delta t\right| < \tau$ \\
$S_2$               & Background & $\left| \Delta \right| > \tau$ \\
$\chi$              & H1L1-correlation in (H1,L1)-spectrograms along a track & Eq.\ref{EQN_c} \\
$\eta$              & Cluster size  & Eq. \ref{EQN_eta}, 0-100\%\\
H1,L1-spectrogram   & Individual detector spectrograms & \S4, Fig. \ref{figTSPDF}\\
$[t_s]$             & $t_{s,{\rm H1}}-t_{s,{\rm L1}}$ & Eq. \ref{EQN_dts4}, Fig. 5\\
$[\tau_s]$          & $\tau_{s,{\rm H1}}-\tau_{s,{\rm L1}}$ & Fig. 5\\
\hline
{\sc $\chi$-image analysis} & & \S1-4\\
\hline
$\chi$              & Correlation along tracks in (merged) spectrograms & Eq. \ref{EQN_c} \\
$\hat{\chi}$        & $\chi(t_s,\tau_s)$ maximal over $f_s,f_0$ & \\
$(t_s^*,\tau_s^*)$  & Location maximum 
$\hat{\chi}^* = \hat{\chi}(t_s^*,\tau_s^*)$ & \\
\hline
Statistics & & \S3-4\\
\hline
PDF                 & Probability density function & \S1 \\
PFA                 & Probability of false alarm & \S1 \\
$T$                 & Duration of data analyzed & Table I \\ 
FAR                 & False Alarm Rate & Table I\\
$p_1$               & PFA$_1$ & $C_1$, Eq. \ref{EQN_C1}, Appendix A\\ 
$p_2$               & PFA$_2$ & $C_2$, Eq. \ref{EQN_C2}\\
\hline
{\sc Computing} & & \\
\hline
Heterogeneous       & Mixed CPU-GPU & Fig. \ref{figHTC2} \\
OpenCL              & Open Compute Language & \cite{amd22,khr22}\\
Local Memory        & Local to a Compute Unit, shared within a work-group & small, low latency\\
Global Memory &  Visible to all work-groups, allocatable by host via PCIe & large, high latency\\
Throughput$^{-1}$   & Wall clock time[s]/(Data length[s]$\times$Template bank size[M]) & Eq. \ref{EQN_HTC1}\\
\hline 
\end{tabular}
}
\end{center}
\label{T3}
\end{table*}

The following gives a brief description of the pipeline to probe the merger sequence Eq. \ref{EQN_MS} optimized for high throughput computation (HTC) to facilitate the analysis in \S3-4 by modern heterogeneous computing. Table \ref{T3} lists some of the relevant symbols.

\subsection{Exascale compute requirements}

To illustrate computational requirements, consider the snippet of 2048\,s of H1L1-data partitioned over $64$ segments of 32\,s duration.
A single spectrogram is produced by $64\times 2^{19}\simeq 3.4\times 10^7$ (H1,L1)-correlations over a bank with $2^{19}=524288$ templates. Evaluated by FFT with $n=2^{17}$ samples per segment at 4096\,Hz, this comprises about 373 teraFLOPs (floating point operations) in single precision (SP, f32). Extended over 161 time-slides $\Delta t$ in $S_{0-2}$ (\S3) is aggregated to 60 petaFLOPs.

The response curves in Fig. \ref{figRes} are extracted from a data-base of 56 injection experiments produced in about $56\times60\simeq 3.3$ exaFLOPs. A similar computational effort holds for $\chi$-image analysis applied to relatively dense scatter plots of spectrograms.

An analysis was carried out on a heterogeneous platform of multi-GPU nodes with synaptic parallel processing (below) over multi-LANs for dynamical load balancing, realizing the required high throughput computations (HTC). 

\subsection{Computing steps}

The two-level pipeline is implemented by heterogeneous computing:
\begin{enumerate}
\item {\em Pre-processing:} H1 and L1 data-frames from the LIGO Open Science Center (LOSC) are paired by GPS start-time, whitened and stored as complex numbers in (H1,L1)-data in single precision in Fortran binary format (little-endian).
\item {\em Whitening:} applied to remove numerous violin modes in the LIGO data. This can be effectively carried out in the Fourier domain by normalizing the spectrum of strain data over intervals of relatively modest bandwidth $B$ (Appendix B, Paper I);
\item {\em Spectrograms:} generated on the basis of (H1,L1) data via butterfly matched filtering is implemented in OpenCL ({\em clButterfly}) on multi-GPU nodes with high bandwidth memory (HBM) for fast evaluation in the Fourier domain, using of pre- and post-callback functions to circumvent PCIe bandwidth \citep{van17};
\item  {\em Rendering:} applied to plot spectrograms in MatLab;
\item  {\em Parameter estimation: } $\chi$-image 
analysis\footnote{van Putten, M.H.P.M., 2018, https://zenodo.org/record/1217028}
of both merged and individal H1 and L1-spectrograms is implemented in OpenCL ({\em clChi}).
\item {\em Post-processing:} gathering results into database for statistical analysis is carried out on CPUs in MatLab.
\end{enumerate}

In \S3, the above is iterated over time-slides ($S_{0-2}$). 
In \S4, iteration is over $n$ template seeds and $m$ offsets by embarrassingly parallel computing. 

\subsection{Limits on performance}

Speed-up by off-loading tasks to GPUs depends crucially on arithmetic intensity. Matched filtering by correlations carried out in the Fourier domain tend to be low arithmetic intensity for arrays of relatively large size, when Global Memory calls are required for matrix transpose. Performance is then well below theoretical peak performance effectively limited by memory bandwidth (Fig. \ref{figHTC2}), pointing to the need for high bandwidth memory (HBM) and a platform comprising multiple multi-GPU nodes. Further optimization obtains by circumventing bandwidth limitations of the peripheral computer interface express (PCIe) with pre- and post-callback functions in the OpenCL implementation clFFT \citep{van17}. 

In contrast, {\em clChi} can be implemented in Local Memory, effectively realizing the computing limit performance on the same GPUs.
We note, however, that scanning over $(\tau_s,f_s,f_0)$ for each step in $t_s$ (about $2\times10^9$ parameter values per data-frame 
of 4096\,s) represents multiple instruction single data (MISD) processing, rather than SIMD in conventional graphics processing. 
Such requires care to avoid overheating the memory bus to Global Memory, by (pseudo-)randomization over $t_s$ in distributing work 
{\em Items} over the large number of stream processors in the GPU. Programs {\em clButterfly} and {\em clChi} are written in C++/F90 with kernels in C99.

\subsection{Multi-GPU node benchmarks}

Benchmarks of butterfly matched filtering can be normalized to clFFT performance\footnote{van Putten, M.H.P.M., 2018, https://zenodo.org/record/1242679}
as follows. With correlations between data and templates evaluated in the Fourier domain, throughput critically depends on the efficiency,
\begin{eqnarray}
\eta_0 = \frac{\mbox{cross-correlation\,rate}}{\mbox{FFT\,transform\,rate}}.
\label{EQN_HTC2}
\end{eqnarray}
Due to limited bandwidth of the Peripheral Computer Interface express (PCIe) between GPU and CPU, there are variations with regard to parallel processing over multi-frame jobs \citep{van17}. It does, however, tend to improve with larger 
job sizes. Net throughput can be expressed as
\begin{eqnarray}
\mbox{throughput}^{-1} = \frac{\mbox{wall\,clock\,time[s]}}{\mbox{data length[s]}\times \mbox{template\,bank\,size[M]}}.
\label{EQN_HTC1}
\end{eqnarray}
Here, the template bank size is expressed in units of 1M templates. 
A throughput in excess of unity defines faster than real-time analysis. 

\begin{figure*}
\includegraphics[scale=0.28]{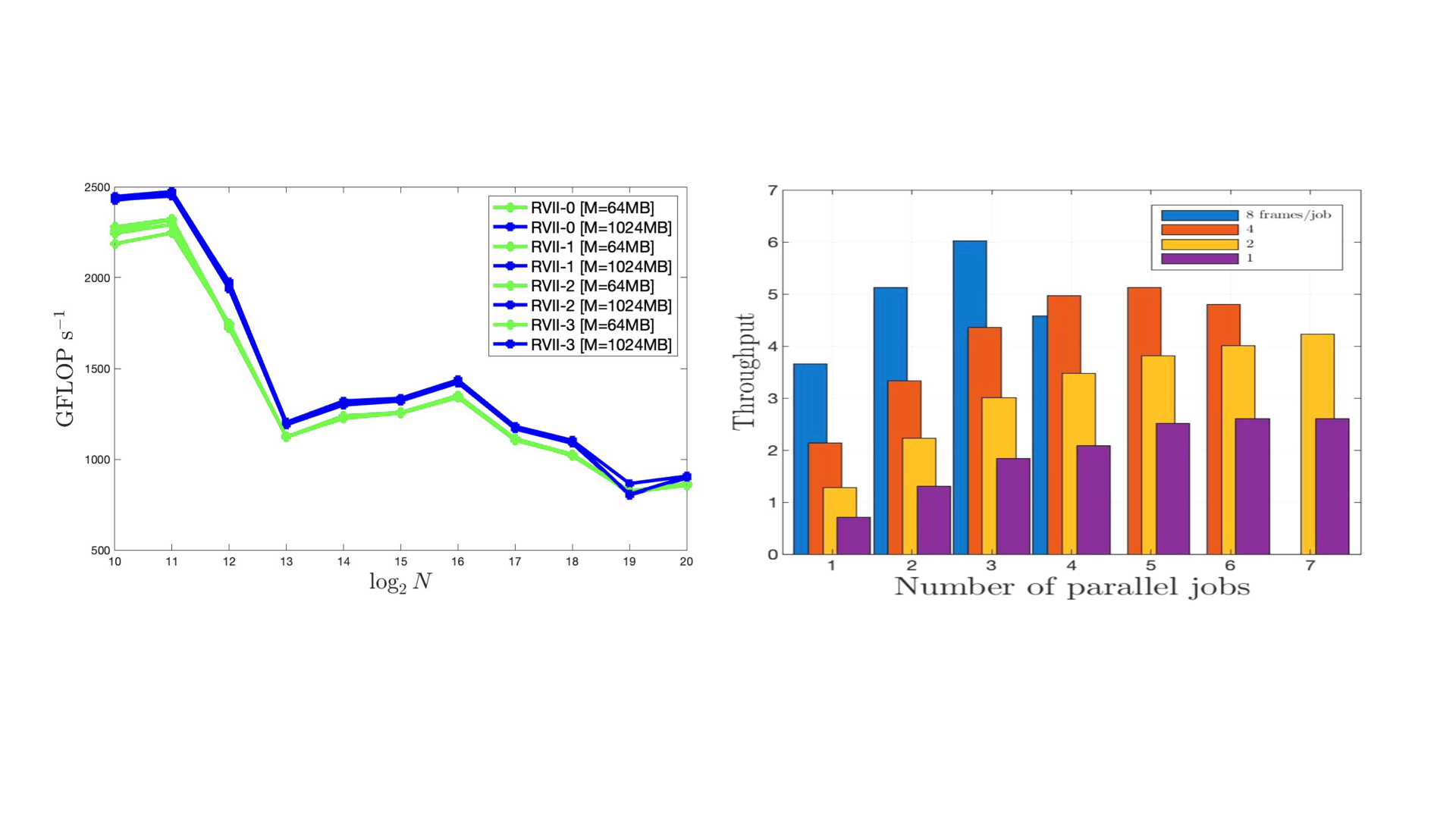}
\vskip-1in
\caption{Performance overview of GPU-accelerated butterfly matched filtering on a quad-GPU node. Performance of clFFT in CSP/Out-of-Place as a function of array size $N$ for two batch sizes measured by allocation $M$ in global memory (left panel). Array sizes exceeding $2^{12}$ cause a drop in performance as matrix transpose exceeds 32\,kB of local memory, limited now by 1TB/s bandwidth to Global Memory in HBM2. We note that {\em clButterfly} uses a default size $N=2^{17}$ (32\,s of data at 4096Hz sampling rate) with batch size over $n=128$ per frame of 4096\,s.
A quad-GPU node hereby features a maximum of about 5 teraFLOPs, circumventing limitations of PCIe bandwidth optimized by pre- and post-callback functions \citep{van17}. The throughput of this quad-GPU node is shown up to six times real-time over a template bank normalized to size $2^{20}$ (right panel). GPU-CPU communication passes through a post-callback function, limiting data-transfer to tails exceeding a threshold $\kappa=2$ in normalized butterfly matched filtering output, effectively 0.1\% of all matched filtering results. Throughput peaks in multi-frame analysis at a cross-correlation rate of about 200kHz.}
\label{figHTC2}
\end{figure*}

Fig. \ref{figHTC2} illustrates \ref{EQN_HTC1}-\ref{EQN_HTC2} for a quad-GPU configuration of an air-cooled node with AMD Radeon VII GPUs featuring $\eta\simeq 47\%$ and a cross-correlation rate up to 200\,kHz between data and templates over segments of 32\,s. 

\subsection{Load balancing by synaptic processing}

Embarrassingly parallel computing for tasks comprising $clButterfly$ (step 3) or $clChi$ (step 5) in \S3-4 are specified as a line-items for processing on a multi-LAN heterogeneous computing platform.

We used synaptic parallel processing in which line-items are requested by the GPU-nodes, rather than issued by a server. Before execution, these line-items are updated with the URLs for input and output data on the LAN hosting a node. This process ensures maximal throughput on heterogeneous platforms of nodes with different performance characteristics. Synaptic parallel processing is written in bash for Linux and OSX.

\end{appendix}
\end{document}